\documentclass[journal]{IEEEtran}

\usepackage[utf8]{inputenc}
\usepackage[T1]{fontenc}
\usepackage{hyperref}
\usepackage{url}
\usepackage{booktabs}
\usepackage{amsfonts}
\usepackage{nicefrac}
\usepackage{microtype}
\usepackage{xcolor}
\usepackage{tikz}
\usepackage{amsmath,amsfonts}
\usepackage{amsthm}
\usepackage{filecontents}
\usepackage{authblk}

\usepackage{nicefrac}
\usepackage{siunitx}
\usepackage{array,framed}
\usepackage{booktabs}
\usepackage{
  color,
  float,
  epsfig,
  wrapfig,
  graphics,
  graphicx,
  subcaption
}
\usepackage{textcomp}
\usepackage{setspace}
\usepackage{latexsym,fancyhdr,url}
\usepackage{enumerate}
\usepackage[ruled,noend,vlined]{algorithm2e}
\usepackage[noend]{algcompatible}
\usepackage{algpseudocode}
\usepackage{graphics}
\usepackage{xparse}
\usepackage{xspace}
\usepackage{multirow}
\usepackage{tabularx}
\usepackage{csvsimple}
\usepackage{balance}
\usepackage{makecell}
\usepackage{colortbl}
\usepackage{enumitem,kantlipsum}
\usepackage{wrapfig}
\usepackage{amsmath}
\usepackage{amssymb}

\SetAlCapNameFnt{\small}
\SetAlCapFnt{\small}

\newcommand*\circled[1]{\tikz[baseline=(char.base)]{\node[shape=circle,draw,inner sep=0.4pt] (char) {#1};}}

\newenvironment{packeditemize}{
	\begin{list}{$\bullet$}{
			\setlength{\labelwidth}{8pt}
			\setlength{\itemsep}{0pt}
			\setlength{\leftmargin}{\labelwidth}
			\addtolength{\leftmargin}{\labelsep}
			\setlength{\parindent}{0pt}
			\setlength{\listparindent}{\parindent}
			\setlength{\parsep}{0pt}
			\setlength{\topsep}{3pt}}}{\end{list}}

\newcommand{\bheading}[1]{{\noindent{\textbf{#1}}}}

\newtheoremstyle{mystyle}%                % Name
  {}%                                     % Space above
  {}%                                     % Space below
  {\itshape}%                             % Body font
  {}%                                     % Indent amount
  {\bfseries}%                            % Theorem head font
  {.}%                                    % Punctuation after theorem head
  { }%                                    % Space after theorem head, ' ', or \newline
  {}%                                     % Theorem head spec (can be left empty, meaning `normal')

\theoremstyle{mystyle}

\newtheorem{theorem}{Theorem}
\newtheorem{definition}{Definition}
\newtheorem{assumption}{Assumption}

\begin{document}

\title{Ownership Verification of DNN Architectures via Hardware Cache Side Channels}
\author[$\dag$]{Xiaoxuan Lou}
\author[$\ddag$]{Shangwei Guo}
\author[$^*$]{Jiwei Li}
\author[$\dag$]{Tianwei Zhang}
\affil[$\dag$]{\textit{School of Computer Science and Engineering, Nanyang Technological University}}
\affil[$\ddag$]{\textit{College of Computer Science, Chongqing University}}
\affil[$^*$]{\textit{Shannon.AI and Zhejiang University} \protect\\
\texttt{\{xiaoxuan001,tianwei.zhang\}@ntu.edu.sg, swguo@cqu.edu.cn, jiwei\_li@shannonai.com}}

% \author{Xiaoxuan Lou,
%         Shangwei Guo,
%         Jiwei Li,
%         and Tianwei Zhang,~\IEEEmembership{Member,~IEEE}
%         %
%         % and Yang Liu,~\IEEEmembership{Senior~Member,~IEEE}
% \thanks{S. Guo is the corresponding author with the  College of Computer Science, Chongqing University, China. E-mail: swguo@cqu.edu.cn}
% \thanks{X. Lou and T. Zhang are with the School of Computer Science and Engineering, Nanyang Technological University, Singapore. E-mail: xiaoxuan001@ntu.edu.sg, tianwei.zhang@ntu.edu.sg}
% \thanks{J. Li is with Shannon.AI and Zhejiang University, China. E-mail: jiwei\_li@shannonai.com}
% % \thanks{Y. Zhang is with the Department of Computer Science and Engineering, Southern University of Science and Technology, China. E-mail: yinqianz@acm.org}
% }

\maketitle

\begin{abstract}
Deep Neural Networks (DNN) are gaining higher commercial values in computer vision applications, e.g., image classification, video analytics, etc. This calls for urgent demands of the intellectual property (IP) protection of DNN models. In this paper, we present a novel watermarking scheme to achieve the ownership verification of DNN architectures. Existing works all embedded watermarks into the model parameters while treating the architecture as public property. These solutions were proven to be vulnerable by an adversary to detect or remove the watermarks. In contrast, we claim the model architectures as an important IP for model owners, and propose to implant watermarks into the architectures. We design new algorithms based on Neural Architecture Search (NAS) to generate watermarked architectures, which are unique enough to represent the ownership, while maintaining high model usability. Such watermarks can be extracted via side-channel-based model extraction techniques with high fidelity.
We conduct comprehensive experiments on watermarked CNN models for image classification tasks and the experimental results show our scheme has negligible impact on the model performance, and exhibits strong robustness against various model transformations and adaptive attacks.
%Compared to prior works, our scheme also enables the use of just one key (watermark) to protect multiple DNN models of various tasks, algorithms and datasets, which can significantly ease the IP management of deep learning assets.

%the first DNN watermarking scheme that aims to model architectures instead of parameters. Since the great profit it brings and the tremendous computational resource it requires, a top-performing DNN model is normally taken as the core Intellectual Property (IP), especially the novel DNN architectures, which has been the bottleneck for continuing the state-of-the-art performance. Unfortunately, previous watermarking methods only focus on the model parameters and are not practically robust. We creatively utilize the Neural Architecture Search (NAS) technology to generate unique but high-usable DNN models, whose architecture can serve as the evidence of the ownership. Besides, we first use the cache side channels to extract and verify the watermark, and we also give a comprehensive analysis on the common operations used in NAS models. The analysis can be generalized to conventional DNN models. We show experimentally that our watermark scheme has negligible impact on the model performance, and is robust against model transformation attacks. In particular, our scheme first enables the IP protection of multiple models with the same key.
\end{abstract}


\IEEEpeerreviewmaketitle

\section{Introduction}
Deep Neural Networks (DNNs) have shown tremendous progress to solve artificial intelligence tasks. Novel DNN algorithms and models were introduced to interpret and understand the open world with higher automation and accuracy, such as image processing \cite{krizhevsky2012imagenet,he2016deep,guo2021topology}, video processing \cite{wang2021neural,li2020adaptive}, natural language processing \cite{devlin2018bert,brown2020language}, bioinformatics \cite{senior2020improved}. With the increased complexity and demand of the tasks, it is more costly to generate a state-of-the-art DNN model: design of the model architecture and algorithm requires human efforts and expertise; training a model with satisfactory performance needs a large amount of computation resources and valuable data samples. Hence, commercialization of the deep learning technology has made DNN models the core Intellectual Property (IP) of AI products and applications.
% 
%made substantial progress, with being applied in tremendous fields. These continuous prosperity mainly benefits from the constant emergence of expert-crafted DNN architectures, such as AlexNet \cite{krizhevsky2012imagenet}, VGGNet \cite{simonyan2014very}, ResNet \cite{he2016deep}, and BERT \cite{devlin2018bert}. In light of such huge success, the community has realized that the model architecture is the bottleneck to continue outperforming state-of-the-art results. A better hand-crafted architectures requires experts to search from the infinite possible choices based on their experience, which however is expensive, time-consuming, and even infeasible for small AI companies. To address this issue, Neural Architecture Search (NAS) is proposed to automatically discover the optimal architectures for training datasets. Starting from the seminal work \cite{zoph2016neural} using reinforcement learning for searching competitive neural architecture, numerous successive advancements \cite{zoph2018learning,pham2018efficient,liu2017hierarchical,brock2017smash,liu2018darts,dong2019searching} have been proposed, and the searched architectures have achieved impressive performance on various tasks, like image classification and natural language processing.
% 
% IP protection problems

Release of DNN models can incur illegitimate plagiarism, unauthorized distribution or reproduction. Therefore, it is of great importance to protect the IP of such valuable assets.
%Even a model is packed in a black-box manner, a malicious entity can still steal the model details from the inference results \cite{tramer2016stealing,wang2018stealing,yu2020cloudleak,jagielski2020high} or execution behaviors \cite{yan2020cache,batina2019csi,hu2020deepsniffer}.
Similar to image watermarking \cite{luo2009reversible,roy2019toward,fang2020deep,li2021concealed,xiong2021robust,you2021truncated,peng2021general}, one common approach for IP protection of DNN models is DNN watermarking, which processes the protected model in a \textit{unique} way such that its owner can recognize the ownership of his model. Existing solutions all implanted the watermarks into the parameters for ownership verification \cite{uchida2017embedding,rouhani2019deepsigns,adi2018turning,zhang2020model,chen2021temporal,wu2020watermarking}. The watermark also needs to guarantee satisfactory performance for the protected model. For example, Adi et al. \cite{adi2018turning} embedded backdoor images with certain trigger patterns into image classification models for IP protection.

Unfortunately, those parameter-based watermarking solutions are not practically robust. An adversary can easily defeat them without any knowledge of the adopted watermarks. First, since these schemes modify the parameters to embed watermarks, the adversary can also modify the parameters of a stolen model to remove the watermarks. Past works have designed such watermark removal attacks, which leverage model fine-tuning \cite{chen2019refit,shafieinejad2019robustness,liu2020removing} or input transformation \cite{guo2020hidden} to successfully invalidate existing watermark methods. Second, watermarked models need to give unique behaviors, which inevitably make them detectable by the adversary. Some works \cite{namba2019robust,aiken2020neural} introduced attacks to detect the verification samples and then manipulate the verification results.

Motivated by the above limitations, we propose a fundamentally different watermarking scheme. Instead of protecting the parameters, we treat the network architecture as the IP of the model. There are a couple of incentives for the adversary to plagiarize the architectures \cite{yan2020cache,hong20200wn}. First, it is costly to craft a qualified architecture for a given task. Architecture design and testing require lots of valuable human expertise and experience. Automated Machine Learning (AutoML) is introduced to search for architectures \cite{zoph2016neural}, which still needs a large amount of time, computing resources and data samples. Second, the network architecture is critical in determining the model performance. The adversary can steal an architecture and apply it to multiple tasks with different datasets, significantly improving the financial benefit. In short, ``the industry considers top-performing architectures as intellectual property'' \cite{hong20200wn}, and ``obtaining them often has high commercial value'' \cite{yan2020cache}.
\textit{Therefore, it is worthwhile to treat the architecture design as an important IP and provide particular protection to it.}

We aim to design a methodology to generate unique network architectures for the owners, which can serve as the evidence of ownership. This scheme is more robust than previous solutions, as maliciously refining the parameters cannot tamper with the watermarks. The adversary can only \textit{remarkably} change the network architecture with large amounts of resources and effort in order to erase the watermarks. This will not violate the copyright, since the new architecture is totally different from the original one, and can be legally regarded as the adversary's own IP. Two questions need to be answered in order to establish this scheme: \emph{(1) how to systematically design architectures, that are unique for watermarking and maintain high usability for the tasks? (2) how to extract the architecture of the suspicious model, and verify the ownership?}

We introduce a set of techniques to address these questions. 
For the first question, we leverage Neural Architecture Search (NAS) \cite{zoph2016neural}. NAS is a very popular AutoML approach, which can automatically discover a good network architecture for a given task and dataset. A quantity of methods \cite{zoph2018learning,pham2018efficient,liu2017hierarchical,brock2017smash,liu2018darts,dong2019searching} have been proposed to improve the search effectiveness and efficiency, and the searched architectures can significantly outperform the ones hand-crafted by humans. Inspired by this technology, we design a novel NAS algorithm, which fixes certain connections with specific operations in the search space, determined by the owner-specific watermark. Then we search for the rest connections/operations to produce a high-quality network architecture. This architecture is unique enough to represent the ownership of the model (Section \ref{sec_nas}).

The second question is solved by cache side-channel analysis. Side-channel attacks are a common strategy to recover confidential information from the victim system without direct access permissions. Recent works designed novel attacks to steal DNN models \cite{yan2020cache,batina2019csi,hu2020deepsniffer}. Our scheme applies such analysis for IP protection, rather than confidentiality breach. The model owner can use side-channel techniques to extract the architecture of a black-box model to verify the ownership, even the model is encrypted or isolated. It is difficult to directly extend prior solutions \cite{yan2020cache,hong2018security} to our scenario, because they are designed only for conventional DNN models, but fail to recover new operations in NAS. We devise a more comprehensive method to identify the types and hyper-parameters of these new operations from a side-channel pattern. 
%recover finer-grained information from NAS models. 
%We use cache side channels to monitor the GEMM (Generalized Matrix Multiply) in the low-level BLAS library (for CNN models) and the activation functions in the high-level DNN framework (for RNN models). 
This enables us to precisely extract the watermark from the target model (Section \ref{sec_sidechannel}).

The integration of these techniques leads to the design of our watermarking framework. Experiments on DNN models for image classification show that our method is immune to common model parameter transformations (fine-tuning, pruning), which could compromise prior solutions. Furthermore, we test new adaptive attacks that moderately refine the architectures (e.g., shuffling operation order, adding useless operations), and confirm their incapability of removing the watermarks from the target architecture. In sum, we make the following contributions:

\begin{packeditemize}
  \item It is the \textit{first} work to protect the IP of DNN architectures. It creatively uses the NAS technology to embed watermarks into the model architectures. 
  \item It presents the \textit{first} positive use of cache side channels to extract and verify watermarks.
  \item It gives a comprehensive side-channel analysis about sophisticated DNN operations that are not analyzed before.
\end{packeditemize}

\section{Background} \label{sec_pre}
\subsection{Neural Architecture Search}
\begin{figure}[t]
  \centering
  \begin{subfigure}{0.46\linewidth}
    \centering
     \includegraphics[width=\linewidth]{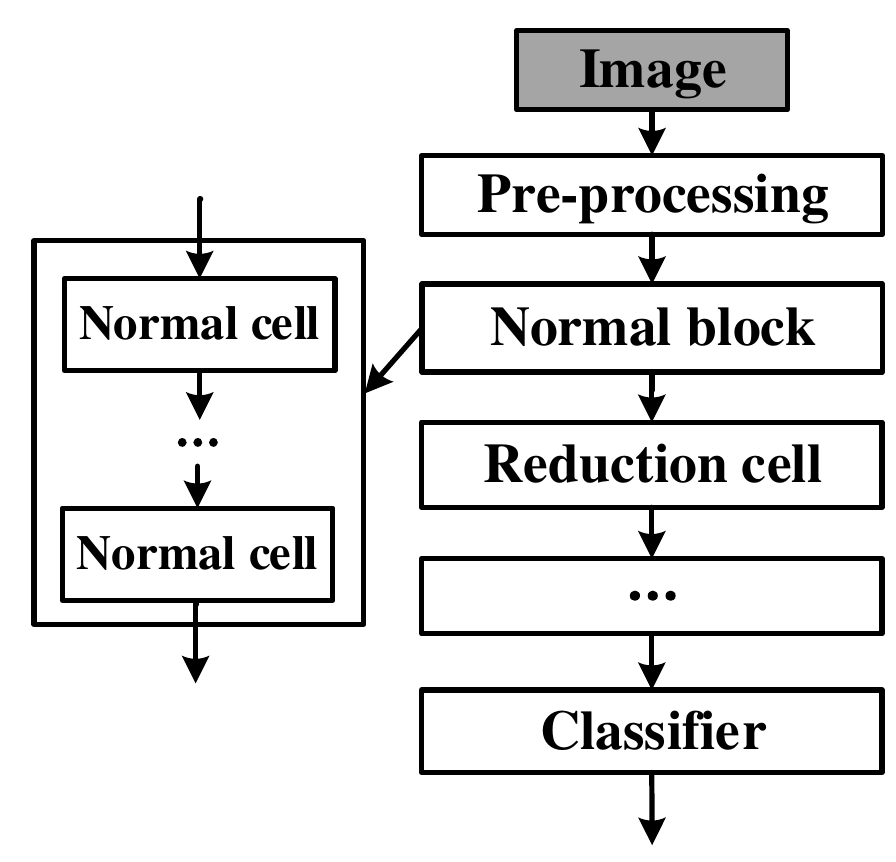}
  \vspace{-5pt}
    \caption{CNN macro-architecture}
    \label{fig_modelarch}
  \end{subfigure}%
  \hspace*{10pt}
  \begin{subfigure}{0.5\linewidth}
    \centering
    \includegraphics[width=\linewidth]{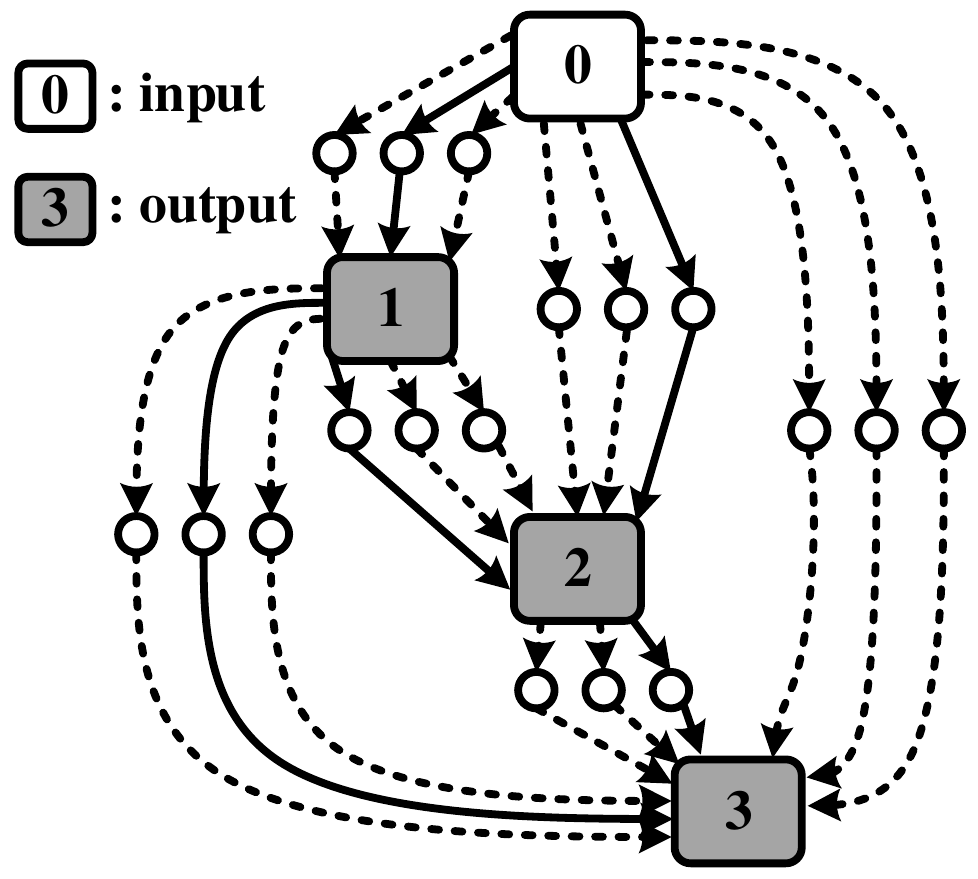}
  \vspace{-5pt}
    \caption{A toy cell \emph{supernet}}
    \label{fig_supernet}
  \end{subfigure}
  \vspace{-5pt}
  \caption{Architecture of a NAS model based on cells}
  \vspace{-10pt}
  \label{fig_nas}
\end{figure}

% Recently Automated Machine Learning (AutoML) has gained huge popularity, due to 
% its capability of building machine learning pipelines and solutions with high efficiency and automation. One of the most popular AutoML techniques is 
NAS \cite{zoph2016neural,elsken2019neural} has gained popularity in recent years, due to its capability of building machine learning pipelines with high efficiency and automation. It systematically searches for good network architectures for a given task and dataset. Its effectiveness is mainly determined by two factors:

\vspace{3pt}
\noindent\textbf{Search space.}
This defines the scope of neural networks to be designed and optimized. Instead of searching for the entire network, a practical strategy is to decompose the target neural network into multiple \emph{cells}, and search for the optimal structure of a cell \cite{zoph2018learning}. Then cells with the identified architecture are stacked in predefined ways to construct the final DNN models. Figure \ref{fig_modelarch} shows the typical architecture of a CNN model based on the popular NAS-Bench-201 \cite{dong2020bench}. It has two types of cells: a \emph{normal cell} is used to interpret the features and a \emph{reduction cell} is used to reduce the spatial size.
%and double the number of filters to roughly maintain the dimension of hidden states. 
A block is composed of several normal cells, and connected to a reduction cell alternatively to form the model. 
%For the RNN model, only the recurrent cell is searched. 

% \begin{figure}[t]
% \centering
% \includegraphics[width=\linewidth]{./fig/fig_modelarch.pdf}
% \caption{Structure of a NAS model based on cells}
% \label{fig_modelarch}
% \end{figure}

% \begin{figure}[h]
%   \centering
%   \includegraphics[width=0.6\linewidth]{./fig/fig_supernet.pdf}
%   \caption{A toy cell \emph{supernet}. The grey squares denote computational nodes, the circles denote candidate operations, and the solid arrows denote chosen connection edges.}
%   \label{fig_supernet}
% \end{figure}

A cell is generally represented as a directed acyclic graph (DAG), where each edge is associated with an operation selected from a predefined operation set \cite{pham2018efficient}. Figure \ref{fig_supernet} gives a toy cell \emph{supernet} that contains four computation nodes (squares) and a set of three  candidate operations (circles). The solid arrows denote the actual connection edges chosen by the NAS method. Such \emph{supernet} enables the sharing of network parameters and avoids unnecessary repetitive training for selected architectures. This significantly reduces the cost of performance estimation and accelerates the search process, and is widely adopted in recent methods \cite{liu2018darts,dong2019searching,chu2019fairnas,bender2018understanding}. 

\vspace{3pt}
\noindent\textbf{Search strategy.}
This defines the approach to seek for good architectures in the search space. Different types of strategies have been designed to enhance the search efficiency and results, based on reinforcement learning \cite{zoph2016neural, zoph2018learning,pham2018efficient}, evolutionary algorithm \cite{real2019regularized,elsken2018efficient} or gradient-based optimization \cite{liu2018darts,chu2020fair,dong2019searching}. Our watermarking scheme is general and independent of the search strategies. 

% $\mathfrak{A} = \texttt{NAS}(\mathbb{S}, \mathcal{D})$, $f=\texttt{retrain}(\mathfrak{A}, \widehat{\mathcal{D}})$
% \shangwei{introduce the NAS and retrain functions}

\subsection{Cache Side Channels}
CPU caches are introduced between the CPU cores and main memory to accelerate the memory access. 
% They store recently used data and instructions for the processor to fetch directly without visiting the main memory. 
Two micro-architectural features of caches enable an adversarial program to perform side-channel attacks and infer secrets from a victim program, even their logical memory is isolated by the operating system. First, multiple programs can share the same CPU cache, and they have contention on the usage of cache lines. Second, the timing difference between a cache hit (fast) and a cache miss (slow) can reveal the access history of the memory lines. As a result, an adversary can carefully craft interference with the victim program sharing the same cache, and measure the access time to infer the victim's access trace. 

A quantity of techniques have been designed over the past decades to realize cache side-channel attacks. Two representative attacks are described as below. (1) In a \textsc{Prime-Probe} attack \cite{liu2015last}, the adversary first fills up the critical cache sets with its own memory lines. Then the victim executes %for a period of time 
and potentially
% touches some critical sets to 
evicts the adversary's data out of the cache. After that, the adversary measures the access time of each memory line loaded previously. A longer access time indicates that the victim used the corresponding cache set. (2) A \textsc{Flush-Reload} attack \cite{yarom2014flush+} requires the adversary to share the critical memory lines with the victim, e.g., via shared library. The adversary first evicts these memory lines out of the cache using dedicated instructions (e.g., \emph{clflush}). After a period of time, it reloads the lines into the cache and measures the access time. A shorter time indicates the lines were accessed by the victim. %The adversary can repeat the above procedure to disclose the victim's memory access trace.

\subsection{Threat Model}
\label{sec:threat-model}

We consider that a model owner designs an architecture using a conventional NAS method, and trains a production-level DNN model $M$. An adversary may obtain an illegal copy of $M$ and use it for profit without authorization. The goal of the model owner is to detect whether a suspicious model $M'$ plagiarizes the architecture from $M$. He has black-box access to the target model $M'$, without any knowledge about the architecture, parameters, training algorithms and hyper-parameters. We consider two sorts of techniques                     an adversary may employ to hide the evidence of architecture plagiarism. (1) \textit{Parameter modification}: the adversary may alter the model parameters (e.g., fine-tuning, model compression, transfer learning) while maintaining similar performance. (2) \textit{Architecture modification}: the adversary may moderately obfuscate the model architecture by changing the execution behaviors of model inference (e.g., reordering the operations, adding useless computations or neurons).  However, we do not consider the case that the adversary redesigns the model architecture completely (e.g., knowledge distillation \cite{ba2013deep,hinton2015distilling}), since the new model architecture is totally different, and can be legally regarded as the adversary's own asset. 

% We further assume the model owner is able to deploy his watermark extraction program on the same physical machine as the target model. 
We further follow the same assumption in \cite{yan2020cache,batina2019csi,hu2020deepsniffer}
that the model owner can extract the inference execution trace of the target model $M'$ via cache side channels. This is applied to the scenario where the suspicious application is securely packed with countermeasures against reverse-engineering, e.g., encryption. For instance, Trusted Execution Environment (TEE), e.g., Intel SGX \cite{mckeen2013innovative} and AMD SEV \cite{amd-sev}, introduces new hardware extensions to provide execution isolation and memory encryption for user-space applications. However, an adversary can exploit TEE to hide their malicious activities, such as side-channel attacks \cite{schwarz2017malware}, rowhammer attacks \cite{gruss2018another} and malware \cite{schwarz2019practical,blackhat18}. Similarly, an adversary can hide the stolen model in TEE when distributing it to the public, so the model owner cannot introspect into the DNN model to obtain the evidence of ownership. With our solution, the model owner can extract watermarks from the isolated enclaves. Note it has been quite common to adopt side-channel techniques to monitor the activities inside TEE enclaves for security purposes \cite{brasser2017software,gotzfried2017cache,hahnel2017high,zhou2022smile}, which is both technically and legitimately allowed.

\section{Related Works} \label{sec_related}
\subsection{DNN Watermarking}
%\subsubsection{Watermarking for DNN}
%Watermarking for DNN has raised substantial concern and 
Numerous watermarking schemes have been proposed for conventional DNN models.
%, where the architectures are publicly available. 
They can be classified into the following two categories:
%and the owners need to train/fine-tune the parameters for better performance on their own databases. We classify these schemes into two categories: parameter-embedding and parameter-poisoning. 

\noindent\textbf{White-box solutions.} 
%Similar to traditional digital watermarking techniques, 
This strategy adopts redundant bits as watermarks and embeds them into the model parameters. 
%The core technique is to design marking algorithms to successfully embed the bits without decreasing the model's performance. 
For instance, Uchida et al. \cite{uchida2017embedding} introduced a parameter regularizer to embed a bit-vector (e.g. signature) into model parameters which can guarantee the performance of the watermarked model. Rouhan et al. \cite{rouhani2019deepsigns} found that implanting watermarks into model parameters directly could affect their static properties (e.g histogram). Thus, they injected watermarks in the probability density function of the activation sets of the DNN layers.  These methods require the owner to have white-box accesses to the model parameters during the watermark extraction and verification phase, which can significantly limit the possible usage scenarios.

\noindent\textbf{Black-box solutions.} This strategy takes a set of unique sample-label pairs as watermarks and embeds their correlation into DNN models. 
%To preserve the functionality and robustness of the watermarked models, the essential component is the generation of carefully crafted watermark samples. 
For examples, Le et al. \cite{le2019adversarial} adopted adversarial examples near the frontiers as watermarks to identify the ownership of DNN models. Zhang et al. \cite{zhang2018protecting} and Adi et al. \cite{adi2018turning} employed backdoor attack techniques to embed backdoor samples with certain trigger patterns into DNN models. Namba et al. \cite{namba2019robust} and Li et al. \cite{li2019prove} generated watermark samples that are almost indistinguishable from normal samples to avoid detection by adversaries.

Different from these works, we propose a new watermarking scheme. Instead of modifying the parameters, our approach makes the architecture design as Intellectual Property, and adopts cache side channels for architecture verification. This strategy can defeat all the watermark removal attacks via parameter transformations. 
%Besides, it is more general for various DNN tasks, datasets and algorithms. 
% Chakraborty et al. \cite{chakraborty2020hardware} also proposed a hardware-assisted IP protection that utilizes the hardware root-of-trust, but it also focuses on model parameters and can only defend fine-tuning attacks. 

% \xiaoxuan{Maybe we can add this ref.}\tianwei{Yes, please}
% Hardware-Assisted IP \cite{chakraborty2020hardware}

\subsection{DNN Model Extraction via Side Channels}
%Past works designed side-channel attacks to extract the property of DNN models. 

%With the widely deployed of deep neural networks in industry settings, the top-performing model itself is usually proprietary and exposed to potential adversaries. There are growing number of attacks aiming to extract the novel architecture of DNN model for profit, where side channel methods account for a large part \cite{hu2020deepsniffer}. 

\noindent\textbf{Cache side channels.}
One popular class of model extraction attacks is based on cache side channels, which monitors the cache accesses of the inference program. 
Hong et al. \cite{hong2018security} recovered the architecture attributes by observing the invocations of critical functions in the deep learning frameworks (e.g., Pytorch, TensorFlow). Similar technique is also applied to NAS models \cite{hong20200wn}. However, these attacks are very coarse-grained. They can only identify convolutions without the specific types and hyper-parameters. 
%such method is very coarse-grained, like all types of convolutions would be converted to the calling of the same function $Conv2d$. The details of operation hyper-parameters should be further revealed from the timing side channels. As a result, 
Yan et al. \cite{yan2020cache} proposed Cache Telepathy, which monitors the GEMM calls in the low-level BLAS library. The number of GEMM calls can greatly narrow down the range of DNN hyper-parameters and then reveal the model architecture. Our method extends this technique to NAS models. Our improved solution can recover more sophisticated operations without the prior knowledge of the architecture family, which cannot be achieved in \cite{yan2020cache}.

%which contain more sophisticated operations that are not discussed in \cite{yan2020cache}. Besides, we also remove the constraint of cache telepathy that requires the knowledge of architecture family. 

\noindent\textbf{Other side channels.}
Some works leveraged other side channels to extract DNN models.
%like power, electromagnetic and timing leakages.  
%Wei et al. \cite{wei2018know} analyzed the power traces of an FPGA-based CNN accelerator to recover the input images to the model. 
Batina et al. \cite{batina2019csi} extracted a functionally equivalent model by monitoring the electromagnetic signals of a microprocessor hosting the inference program. 
Duddu et al. \cite{duddu2018stealing} found that models with different depths have different execution time, which can be used as a timing channel to leak the network details. 
Memory side-channels were discovered to infer the network structure of DNN models on GPUs \cite{hu2020deepsniffer} and DNN accelerators \cite{hua2018reverse}. Future work will apply those techniques to our scheme.

\section{Preliminaries} \label{sec_watermark}
% In this section, we formally define the watermarking scheme for NAS systems and theoretically construct a watermarking scheme from a strong stamp an ideal analyzer.

\subsection{Definition of A NAS Method}

% \xiaoxuan{The search space of NAS consist of two parts: network topology and operations used on edges. Consider a NAS cell that contains $\mathcal{B}$ computing nodes, its topology can be denoted as a graph $\mathcal{G} = (\mathcal{N}, \mathcal{E})$, where $\mathcal{N}$ is the set of $\mathcal{B}$ nodes and $\mathcal{E}=\{\mathcal{E}_1, ..., \mathcal{E}_B\}$ is the set of edges. Let $\mathcal{O}$ denote the set of candidate NAS operations, then the search space of a NAS cell is $\mathcal{S}_i = \{\mathcal{G}, \mathcal{O}\}$. Finally we combine the search spaces of all cell as $\mathbb{S}$, from which we look for an optimal architecture $\mathfrak{A}$.}

% With all above techniques, latest NAS methods can significantly reduce the search effort from thousands of GPU days \cite{zoph2016neural, zoph2018learning} to a few hours \cite{liu2018darts,dong2019searching}, and the generated model has competitive prediction accuracy as the state-of-the-art ones. Hence, NAS is definitely a promising direction for deep learning engineering.
In this paper, we mainly focus on NAS methods using the cell-based search space, as it is the most popular and efficient strategy. Formally, we consider a NAS task, which aims to construct a model architecture containing $N$ cells:
$\mathfrak{A} = \{c_1,..., c_N\}$. The search space of each cell is denoted as $S=(\mathcal{G}, \mathcal{O})$. $\mathcal{G} = (\mathcal{N}, \mathcal{E})$ is the DAG representing the cell \emph{supernet}, where set $\mathcal{N}$ contains two inputs $(a, b)$ from previous cells and $\mathcal{B}$ computing nodes in the cell, i.e., $\mathcal{N} = \{a, b, \mathcal{N}_1, ..., \mathcal{N}_\mathcal{B}\}$; $\mathcal{E}=\{\mathcal{E}_1, ..., \mathcal{E}_B\}$ is the set of all possible edges between nodes and $\mathcal{E}_j$ is the set of edges connected to the node $\mathcal{N}_j$ ($1 \leq j \leq B$). Each node can only sum maximal two inputs from previous nodes. $\mathcal{O}$ is the set of candidate operations on these edges. 
% $B$ is the number of computation nodes in the cell; $\mathcal{E}=\{\mathcal{E}_1, ..., \mathcal{E}_B\}$ is the set of all possible edges between nodes and $\mathcal{E}_j$ is the set of edges connected to the $j$-th node ($1 \leq j \leq B$); $\mathcal{O}$ is the set of candidate operations on these edges.
% Hence, the supernet of a cell can be represented as a matrix $M^c$
% whose row size is the size of $\mathcal{E}$ and column size is the size of $\mathcal{O}$. Each element at position $(x,y)$ can be 1 if edge $x$ is attached with operation $y$, or 0 otherwise. 
Then we combine the search spaces of all cells as $\mathbb{S}$, from which we try to look for an optimal architecture $\mathfrak{A}$. The NAS method is defined as below:

% According to Definition \ref{def:nas}, the structure of a NAS model consists of $N$ computational cells: $\mathfrak{A} = \{c_1,..., c_N\}$. We denote the search space of each cell as $(\mathcal{B}, \mathcal{E}, \mathcal{O})$: $\mathcal{B}$ is the number of computational nodes in the cell; $\mathcal{E}=\{\mathcal{E}_1, ..., \mathcal{E_B}\}$ is the set of all possible edges and $\mathcal{E}_j$ is the set of edges connected to the $j$-th node ($1 \leq j \leq |\mathcal{B}|$); $\mathcal{O}$ is the set of candidate operations attached to these edges.
% Since NAS normally requires each node has maximal two inputs from previous nodes, so at most two elements can be selected in $\mathcal{E}_j$.
%This restriction also impacts the generation of marking keys, which will be discussed later.
% For each node $j$, two input edges from $\mathcal{E}_j$ need to be determined and attached with chosen operations.

% \shangwei{merge the two paragraphs above.}

\begin{definition}\label{def:nas}(NAS)
   A NAS method is a machine learning algorithm that iteratively searches optimal cell architectures from the search space $\mathbb{S}$ on the proxy dataset $\mathcal{D}$. These cells construct one architecture $\mathfrak{A} = \{c_1,..., c_N\}$, i.e., $\mathfrak{A} = \emph{\texttt{NAS}}(\mathbb{S}, \mathcal{D}).$
\end{definition}
After the search process, $\mathfrak{A}$ is trained from the scratch on the task dataset $\overline{\mathcal{D}}$ to learn the optimal parameters. The architecture $\mathfrak{A}$ and the corresponding parameters give the final DNN model $f = \texttt{train}(\mathfrak{A}, \overline{\mathcal{D}}).$

% $\mathfrak{A} = \texttt{NAS}(\mathbb{S}, \mathcal{D})$, $f=\texttt{retrain}(\mathfrak{A}, \widehat{\mathcal{D}})$
% \shangwei{introduce the NAS and retrain functions}

\subsection{Definition of A Watermarking Scheme}
\label{sec:wm-def}
% show why we use side channel leakage for verification
A watermarking scheme for NAS enables the ownership verification of DNN models searched from a NAS method. This is formally defined as below:

\begin{definition}\label{def: watermarking}
    A watermarking scheme for NAS is a tuple of probabilistic polynomial time algorithms (\emph{\textbf{WMGen}}, \emph{\textbf{Mark}} \emph{\textbf{Verify}}), where
    \begin{packeditemize}
        \item \emph{\textbf{WMGen}} takes the search space of a NAS method as input and outputs secret marking key $mk$ and verification key $vk$.
        \item \emph{\textbf{Mark}} outputs a watermarked architecture $\mathfrak{A}$, given a NAS method, a proxy dataset $\mathcal{D}$, and $mk$.
        \item \emph{\textbf{Verify}} takes the input of $vk$ and the monitored side-channel trace, and outputs the verification result of the watermark in $\{0,1\}$.
    \end{packeditemize}
\end{definition}

A strong watermarking scheme for NAS should have the following properties \cite{zhang2018protecting,adi2018turning}.
% \shangwei{USENIX18 uses games between attacker and defender to describe the properties. we can follow the paper if needed.}

\bheading{Effectiveness.} The watermarking scheme needs to guarantee the success of the ownership verification over the watermarked $\mathfrak{A}$ using the verification key. Formally, 
\begin{equation}
    Pr[\textbf{Verify}(vk, \mathbb{T}) = 1] = 1,
\end{equation}
where $\mathbb{T}$ is the monitored side-channel trace from $\mathfrak{A}$.

\bheading{Usability.} let $\mathfrak{A}_0$ be the original architecture without watermarks. For any data distribution $\mathcal{D}$,
the watermarked architecture $\mathfrak{A}$ should exhibit competitive performance compared with $\mathfrak{A}_0$ on the data sampled from $\mathcal{D}$, i.e.,
\begin{equation}
    |Pr[f_0(x) = y|(x,y)\thicksim \mathcal{D}] - Pr[f(x) = y|(x,y)\thicksim \mathcal{D}] | \leq \epsilon.
\end{equation}
where $f = \texttt{train}(\mathfrak{A}, \overline{\mathcal{D}})$ and $f_0 = \texttt{train}(\mathfrak{A}_0, \overline{\mathcal{D}})$.

\begin{figure*}[t]
  \centering
  \includegraphics[width=0.8\linewidth]{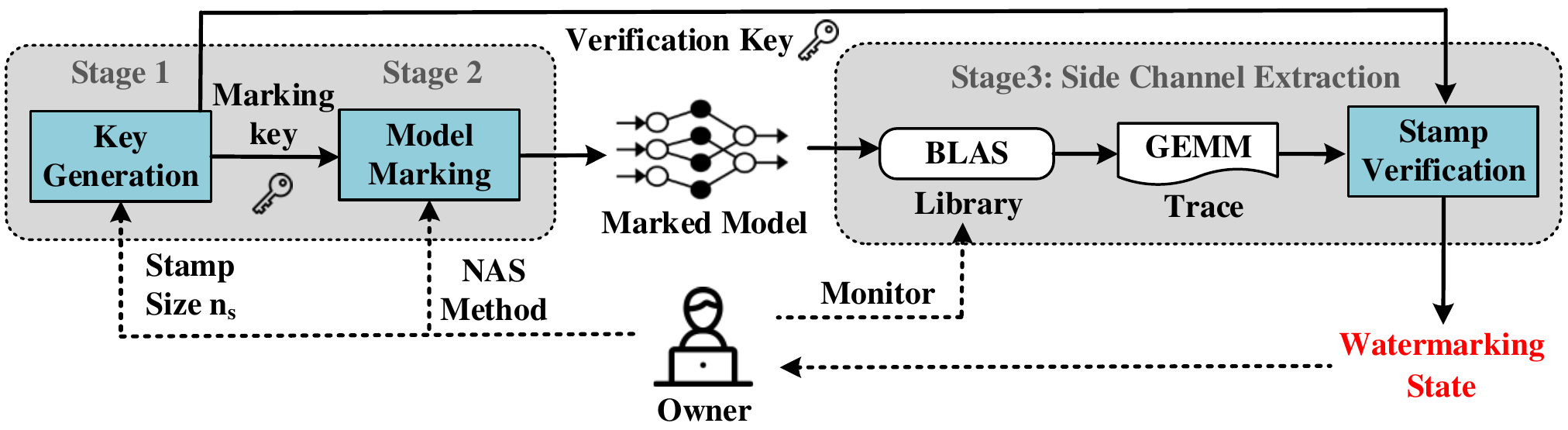}
  \vspace{-5pt}
  \caption{Overview of our watermarking framework}
  \vspace{-5pt}
  \label{fig_intro}
\end{figure*}

\bheading{Robustness.} Since a probabilistic polynomial time adversary may modify $f$ moderately, we expect the watermark remains in $\mathfrak{A}$ after those changes. Formally, let $\mathbb{T}'$ be the side-channel leakage of a model $f'$ transformed from $f$, where $f’$ and $f$ are from the same architecture  $\mathfrak{A}$ with similar performance. We have
%let $\widehat{\mathfrak{A}}' \gets \mathcal{A}(\mathcal{O}^{\mathcal{S}}, \mathcal{A}(\mathcal{O}^{\mathcal{D}}, \widehat{\mathfrak{A}})$

\begin{equation}
    Pr[\textbf{Verify}(vk, \mathbb{T}')=1] \geq 1-\delta
    % \texttt{Compare}(vk', vk) \leq \tau,
\end{equation}
%if $f’$ and $f$ have similar performance.% $\texttt{Compare}$ is a function that calculates the ratio of same operations at the same connected edges between $vk'$ and $vk$.

\bheading{Uniqueness.} A normal user can follow the same NAS method to learn a model from the same proxy dataset. Without the marking key, the probability that this common model contains the same watermark should be smaller than a given threshold $\delta$. Let $\mathbb{T}'$ be the side channel leakage of a common model learned with the same dataset and NAS method,  
% Let $vk'$ be the extracted watermark from the DL model learned with the same training dataset and NAS method. 
we have
\begin{equation}
    Pr[\textbf{Verify}(vk, \mathbb{T}')=1] \leq \delta.
\end{equation}

\section{Our Watermarking Scheme} \label{sec_nas}
%Different from past works, 
% Our solution embeds watermarks into the neural architectures, and uses cache side channels to verify the ownership. Figure \ref{fig_intro} shows the overview of our scheme, which includes watermark generation (Section \ref{sec:key-gen}), embedding (Section \ref{sec_modelmark}) and verification (Section \ref{sec_verifymark}). We introduce a novel algorithm for each stage, followed by a theoretical analysis (Section \ref{sec:theoretic}).
%As a result, existing solutions for generating, embedding and verifying watermarks cannot be extended to our mechanism. This section introduces our novel watermarking mechanism. We describe how we generate a marking key and identify the stamp (Section \ref{sec:key-gen}), embed the stamp into the model (Section \ref{sec_modelmark}), and verify the embedded watermarks (Section \ref{sec_verifymark})

%This section describes the theoretical procedure of conducting a privately verifiable watermark in the NAS model. Fitst, the marking key $mk$ is determined, which configures the cell stamp $\widehat{M}^c$ and reduces the NAS search space. Then the NAS method is invoked to find out the optimal cells from the reduced search space, which are further stacked with certain hyper-parameters (e.g., input channels and size) to construct the DNN model for specific tasks. Combining the embedded stamps and model hyper-parameters, the verification key $vk$ is finally obtained.

Figure \ref{fig_intro} shows the overview of our watermarking framework, which consists of three stages. At stage 1 , the model owner generates a unique watermark and the corresponding key pair $(mk, vk)$ using the algorithm \textbf{WMGen} (Section \ref{sec:key-gen}). At stage 2, he adopts a conventional NAS method with the marking key $mk$ to produce the watermarked architecture following the algorithm \textbf{Mark} (Section \ref{sec_modelmark}). He then trains the model from this architecture. Stage 3 is to verify the ownership of a suspicious model: the owner collects the side-channel information at inference, and identifies any potential watermark based on the verification key $vk$ using the algorithm \textbf{Verify} (Section \ref{sec_verifymark}). Below we describe the details of each stage, followed by a theoretical analysis (Section \ref{sec:theoretic}).

\subsection{Watermark Generation (WMGen)}
\label{sec:key-gen}
% A watermark for a NAS model consists of multiple \textit{stamps}. 
According to Definition \ref{def:nas}, a NAS architecture is a composition of cells. Each NAS cell is actually a sampled sub-graph of the \emph{supernet} $\mathcal{G}$, where the attached operations are identified by the search strategy. To generate a watermark, the model owner selects some edges from $\mathcal{G}$ which can form a path. We select the edges in a path because the executions of their operations have dependency (see the red edges in Figure \ref{fig_cells}). So an adversary cannot remove the watermarks by shuffling the operation order at inference. Then the model owner fixes each of these edges with a randomly chosen operation. The set of the fixed edge-operation pairs $\{s_e : s_o\}$ inside a cell is called a \emph{stamp}, as defined below:

%The ownership verification of NAS models equals to that of cells. Considering the model transformation and adjustment attacks, we select certain edges in $\mathcal{G}$ and fix them with chosen operations for each cell during the search process, where the fixed edge-operation pairs $\{s_e : s_o\}$ should be causal, i.e., edges in $s_e$ should form a consecutive path. The set of the fixed edge-operation pairs inside a cell is called a \emph{stamp}. It is formally defined as below:

% definition of stamp
\begin{definition}\label{def:stamp}(Stamp)
	A stamp for a cell is a set of edge-operation pairs $\{s_e : s_o\}$, %\shangwei{how abut $k_i = \{\mathcal{E}_s : \mathcal{O}_s\}$}, 
	where $s_e$, $s_o$ denote the selected edges in a path and the corresponding operations, respectively.%, and edges in $s_e$ should form a consecutive path.
	%  During the NAS process, the operations of the cell are fixed on the corresponding positions with $\widehat{M}^c$, i.e., for $\forall 0\leq i < m, 0 \leq j < n$, $M^{c}(i,j) = \widehat{M}^c(i,j)$ if $\widehat{M}^c(i,j) \neq 0$.
\end{definition}

The combination of the stamps of all the cells form a watermark for a NAS architecture:

%can succinctly represent the architecture. Inspired by this intuition, we formally define the watermark for a NAS architecture as follows.
\begin{definition}\label{def:watermark}(Watermark)
	Consider a NAS method with a proxy dataset $\mathcal{D}$ and search space $\mathbb{S}$. $\mathfrak{A} = \{c_1,..., c_N\}$ represents the neural architecture produced from this method. A watermark for $\mathfrak{A}$ is a set of stamps ${mk_1,...,mk_N}$, where $mk_i$ is the stamp of cell $c_i$.
\end{definition}

%\begin{wrapfigure}{R}{0.6\linewidth}
%\vspace{-10pt}
%\hspace{0.05\linewidth}
%\begin{minipage}{0.89\linewidth}
\begin{algorithm}[t]
	\caption{Marking Key Generation (\textbf{WMGen})} \label{alg1}
	\KwIn{\# of fixed edges $n_s$, search space $\mathbb{S}=(\mathcal{G}, \mathcal{O})$}
	\KwOut{marking key $mk$, verification key $vk$}
%	Retrieve cell search space $S=(\mathcal{G}, \mathcal{O})$ from $\mathbb{S}$, $\mathcal{G} = (\mathcal{N}, \mathcal{E})$, $\mathcal{N} = \{a, b, \mathcal{N}_1, ..., \mathcal{N}_\mathcal{B}\}$, $\mathcal{E}=\{\mathcal{E}_1, ..., \mathcal{E}_B\}$ \\
%	$S_e = \{ \}$ is a set of edges in a path with length $n_s$ \\
%	$P_1 = \{a, b\} \circ \mathcal{N}_1$  \\
%	\For{i from 2 to $\mathcal{B}$}{
%	    $P_i = (P_{i-1} \cup ... \cup P_1 \cup \{a\} \cup \{b\}) \circ \mathcal{N}_i$ \\
%	}
%	\For{$p$ in $P_\mathcal{B}$}{
%	    \If{$|p| \geq n_s$}{
%	        $S_e = S_e \cup \texttt{GetSubPath}(p, n_s)$ \\
%	    }
%	}
	$S_e = \texttt{GetPath}(\mathcal{G}, n_s)$\\
	\For{ i \emph{from} 1 to N}{
	    $s_e \leftarrow$ randomly select one path from $S_e$ \\
	    $s_o \leftarrow$ randomly select $n_s$ operations from $\mathcal{O}$ for $s_e$\\
	    $mk_i = \{s_e : s_o\}$, $vk_i = s_o$ \\
	}
	\Return $mk$ = ($mk_1$, ..., $mk_N$), $vk$ = ($vk_1$, ..., $vk_N$)
\end{algorithm}
%\end{minipage}
%\vspace{-15pt}
%\end{wrapfigure}

Algorithm \ref{alg1} illustrates the detailed procedure of constructing a watermark and the corresponding marking and verification keys $(mk, vk)$. 
%The operation $\{\emph{set}\} \circ \mathcal{N}_i$ appends the node $\mathcal{N}_i$ to each element in the \emph{set}, generating a set $P_i$ of possible paths from the cell inputs to node $\mathcal{N}_i$. Specifically, $P_\mathcal{B}$ contains all the candidate paths in the cell \emph{supernet} $\mathcal{G}$.
Given the supernet $\mathcal{G}$, we call function \texttt{GetPath} (Algorithm \ref{alg_getpath} in Appendix \ref{sec_path}) to obtain a set $S_e$ of all the possible paths with length $n_s$, where $n_s$ is the predefined number of stamp edges ($1 \leq n_s \leq \mathcal{B}$). Then for each cell $c_i$, we randomly sample a path $s_e$ from $S_e$. Edges in the selected path are attached with fixed operations $s_o$ chosen by the model owner to form the cell stamp $mk_i = \{s_e : s_o\}$. 
% a random operation, to form the cell stamp $mk_i = \{s_e : s_o\}$. 
Finally we can construct a marking key $mk = (mk_1, ..., mk_N)$. The verification key is $vk$ = ($vk_1$, ..., $vk_N$), where $vk_i$ is the fixed operation sequence $s_o$ in cell $c_i$.

In our implementation, we randomly sample the paths and operations for the marking key. It is also possible the model owner crafts the stamps based on his own expertise. He needs to ensure the design is unique and has very small probability to conflict with other models from the same NAS method. We do not discuss this option in this paper.

\subsection{Watermark Embedding (Mark)} \label{sec_modelmark}
To generate a competitive DNN architecture embedded with the watermark, we fix the edges and operations in the marking key $mk$, and apply a conventional NAS method to search for the rest connections and operations for a good architecture. This process will have a smaller search space compared to the original method. However, as shown in previous works \cite{zoph2018learning, liu2018darts}, there are multiple sub-optimal results with comparable performance in the NAS search space, which makes random search also feasible. Hence, we hypothesize that we can still find out qualified results from the reduced search space.  
Evaluations in Section \ref{sec_eva} verify that the reduced search space incurs negligible impact on the model performance.

Algorithm \ref{alg2} shows the procedure of embedding the watermark to a NAS architecture. For each cell $c_i$ in the architecture, we first identify the fixed stamp edges and operations $\{s_e : s_o\}$ from key $mk_i$. Then the cell search space $S$ is updated as $(\mathcal{G} = (\mathcal{N}, \overline{\mathcal{E}}), \mathcal{O})$, where $\overline{\mathcal{E}}$ is the set of connection edges excluding those fixed ones: $\overline{\mathcal{E}}= \mathcal{E} - s_e$. The updated search spaces of all the cells are combined to form the search space $\mathbb{S}$, from which the NAS method is used to find a good architecture $\mathfrak{A}$ containing the desired watermark. 

\noindent\textbf{Discussion}. We describe our watermarking scheme with the NAS technique. It is worth noting that our methodology can also be applied to the hand-crafted architectures. The model owner only needs to inject the stamp edges to some locations inside his designed architecture and then train the model. We consider the evaluation of this strategy as future work.

\begin{algorithm}[t]
	\caption{Watermark Embedding (\textbf{Mark})} \label{alg2}
	\KwIn{marking key $mk$, 
	%model macro-architecture $\mathfrak{A}$, 
	$\texttt{NAS}$ method, proxy dataset $\mathcal{D}$}
	\KwOut{watermarked architecture $\mathfrak{A}$}
	$\mathbb{S} \leftarrow $ search space of the whole model \\
	\For{\emph{each cell} $c_i$}{
	    retrieve $\{s_e : s_o\}$ from $mk_i$ \\
	    $\overline{\mathcal{E}}= \mathcal{E} - s_e$  \\
	    $S=(\mathcal{G} = (\mathcal{N}, \overline{\mathcal{E}}), \mathcal{O})$ \\
	    $\mathbb{S}$.append($S$) \\
	}
	$\mathfrak{A} = \texttt{NAS}(\mathbb{S}, \mathcal{D})$ \\
%	$f=\texttt{train}(\mathfrak{A}, \overline{\mathcal{D}})$ \\
	\Return $\mathfrak{A}$
\end{algorithm}

\subsection{Watermark Verification (Verify)}
\label{sec_verifymark}
During verification, we utilize cache side channels to capture an execution trace $\mathbb{T}$ by monitoring the inference process of the target model $M'$. Details about side-channel extraction can be found in Section \ref{sec_sidechannel}. Due to the existence of extra computations like concatenating and preprocessing, cells in $\mathbb{T}$ are separated by much larger time intervals and can be identified as sequential leakage windows. 
% pre-processing and post-processing operations, the latency between two cells is much longer and human-noticeable. So we can easily divide this trace into sequential windows, with each one representing the pattern of a NAS cell. 
If $\mathbb{T}$ does not have observable windows, we claim it is not generated by a cell-based NAS method and is out of the consideration. A leakage window further contains multiple clusters, each of which corresponds to an operation inside the cell. 

Algorithm~\ref{alg3} describes the verification process. First the side-channel leakage trace $\mathbb{T}$ is divided into cell windows, and for the $i$-th window, we retrieve its stamp operations $s_o$ from $vk_i$. Then the cluster patterns in the window are analyzed in sequence. Since the adversary can possibly shuffle the operation order or add useless computations to obfuscate the trace, \emph{we only verify if the stamp operations exist in the cell in the correct order, which is not affected by the obfuscations due to their execution dependency, while ignoring other operations}. Besides, since the adversary may inject useless cell windows to obfuscate the verification, we only consider cells that contain the expected side-channel patterns and skip other cells. Once the number of verified cells is equal to the size of generated verification key, we can claim the architecture ownership of the DNN model.

%\begin{minipage}{0.48\textwidth}\small
%\vspace{0pt}

%\end{minipage}
%\hfill
%\begin{minipage}{0.48\textwidth}\small
%\vspace{0pt}
\begin{algorithm}[t]
	\caption{Watermark Verification (\textbf{Verify})} \label{alg3}
	\KwIn{verification key $vk$, monitored trace $\mathbb{T}$, \# of fixed edges $n_s$}
	\KwOut{verification result}
	Split $\mathbb{T}$ into \emph{cell windows} \\
	go\_on = 1, verified\_wins = 0 \\
	\For{\emph{each} $window_i$ \emph{in} $\mathbb{T}$}{
	   % stamp edge order $s_{id} \Leftarrow k_i \Leftarrow vk$ \\
	   \If{go\_on == 1}{retrieve $s_o$ from $vk_i$, 
	    \emph{id} $\leftarrow$ 0 \\}
	    \For{\emph{each} $cluster$ \emph{in} $window_i$}{
	        \If{\texttt{\emph{match}}($cluster$, $s_o[id]$) = \emph{True}}{
	                $id += 1$ \\
	        }
	    }
	   % $r = (id = n_s) ? True : False $  \\
	    \If{$id == n_s$}{go\_on = 1, verified\_wins += 1}
	    \Else{go\_on = 0}
	} %\[\]
	\Return (verified\_wins == vk.size()) ? True : False
\end{algorithm}

\subsection{Theoretical Analysis}
\label{sec:theoretic}
We theoretically prove that our algorithms (\textbf{WMGen}, \textbf{Mark}, \textbf{Verify}) 
form a qualified watermarking scheme for NAS architectures. We first assume the search space restricted by the watermark is still large enough for the owner to find a qualified architecture.

\begin{assumption}\label{ass:stamp}
    Let $\mathbb{S}_0$, $\mathbb{S}$ be the search spaces before and after restricting a watermark in a NAS method, $\mathbb{S}_0 \supseteq \mathbb{S}$. $\mathfrak{A}_0 \in \mathbb{S}_0$ is the optimal architecture for an arbitrary data distribution $\mathcal{D}$. $\mathfrak{A}$ is the optimal architecture in $\mathbb{S}$, The model accuracy of $\mathfrak{A}$ is no smaller than that of $\mathfrak{A}_0$ by a relaxation of $\frac{\epsilon}{N}$.
%    For $\forall \ \mathfrak{A}^* \in \mathbb{S}_0\setminus \mathbb{S}$, the accuracy of $\mathfrak{A}^*$ is no larger than that of $\mathfrak{A}$ by a relaxation of $\frac{\epsilon}{N}$. 
    
%    \shangwei{the accuracy of $\mathfrak{A}^* \leq$ the acc of $\mathfrak{A}$ + $\frac{\epsilon}{N}$}
\end{assumption}

%To ensure the success of the verification process, 
We further assume the existence of an ideal analyzer that can recover the watermark from the given side-channel trace.
\begin{assumption}\label{ass:leakage}
    Let  $mk$ and $vk$ be the marking and verification keys of a DNN architecture $\mathfrak{A} = \{c_1,..., c_N\}$. For $\forall$ $mk, vk,$ and $\mathfrak{A}$, there is a leakage analyzer $P$ that is capable of recovering all the stamps of $\{c_i\}_{i=1}^{N}$ from a corresponding cache side-channel trace.
\end{assumption}

%Let $\texttt{NAS}$ be a NAS method that is integrated with our algorithms of watermark generation, embedding and verification. 
With the above two assumptions, we prove the following theorem, and the proof can be found in Appendix \ref{sec:proof}.
%that our algorithms (\textbf{WMGen}, \textbf{Mark}, \textbf{Verify}) can satisfy the properties in Section \ref{sec:wm-def}. The proof can be found in Appendix \ref{sec:proof}.

\begin{theorem}
	With Assumptions \ref{ass:stamp}-\ref{ass:leakage}, Algorithms \ref{alg1}-\ref{alg3} form a watermarking scheme that satisfies the properties of effectiveness, usability, robustness, and uniqueness in Section \ref{sec:wm-def}.
\end{theorem}

%\begin{theorem}
%	The proposed Algorithms \ref{alg1}-\ref{alg3} form a watermarking scheme that satisfies the properties of effectiveness, usability, robustness, and uniqueness when Assumptions \ref{ass:stamp}-\ref{ass:leakage} hold true.
%\end{theorem}

\section{Side Channel Extraction} \label{sec_sidechannel}
% \subsection{Extraction Technique}
Given a suspicious model, we aim to extract the embedded watermark using cache side channels. Past works proposed cache side channel attacks to steal DNN models \cite{yan2020cache,hong2018security}. However, these attacks are only designed for conventional DNN models and cannot extract NAS models with more sophisticated operations (e.g., separable convolutions, dilated-separable convolutions). Besides, the adversary needs to have the knowledge of the target model's architecture family (i.e., the type of each layer), which cannot be obtained in our case. 

We design an improved methodology over Cache Telepathy \cite{yan2020cache} to extract the architecture of NAS models by monitoring the side-channel pattern from the BLAS library.
% \footnote{For RNN models, we monitor the high-level deep learning framework, as the BLAS library does not leak information about the model}. 
We take OpenBLAS as an example, which is a mainstream library for many deep learning frameworks (e.g., Tensorflow, PyTorch). Our method is also generalized to other BLAS libraries, such as Intel MKL. 
We make detailed analysis about the leakage pattern of common operations used in NAS, and describe how to identify the operation type and hyper-parameters.

\begin{figure*}[hbt]
  \centering
  \includegraphics[width=0.85\linewidth]{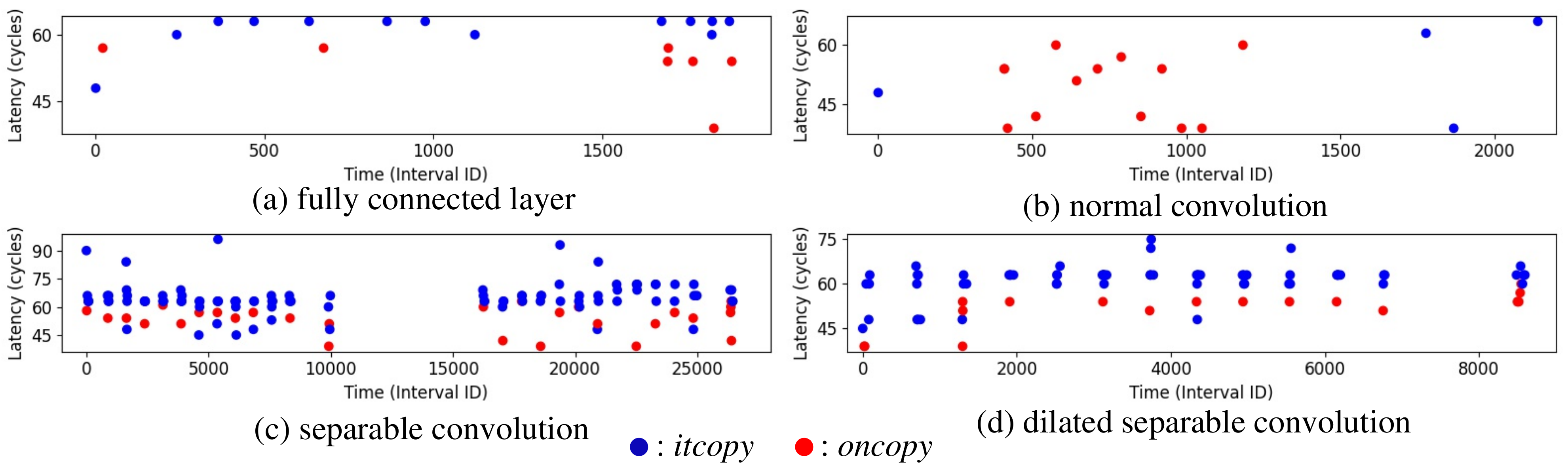}
  \vspace{-10pt}
  \caption{Side-channel patterns of  four operations in NAS.}
  \label{fig_patterns}
  \vspace{-10pt}
\end{figure*}

\subsection{Method Overview}
State-of-the-art NAS algorithms \cite{zoph2018learning, liu2017hierarchical, liu2018darts, dong2020bench} commonly adopt eight classes of operations: (1) identity, (2) fully connected layer, (3) normal convolution, (4) dilated convolution, (5) separable convolution, (6) dilated-separable convolution, (7) pooling and (8) various activation functions. Note that although \emph{zeroize} is also a common operation in NAS, we do not consider it, as it just indicates a lack of connection between two nodes and is not actually used in the search process.

%For the implementation of DNN operations, in particular the complex ones like fully connected layers and convolutions, 
These operations are commonly implemented in two steps. (1) The high-level deep learning framework converts an operation to a matrix multiplication: $C=\alpha A \times B + \beta C$, where input A is an $m \times k$ matrix and B is a $k \times n$ matrix, output C is an $m \times n$ matrix, and both $\alpha$ and $\beta$ are scalars; (2) The low-level BLAS library performs the matrix multiplication with the GEMM algorithm, which divides matrix A and B into smaller ones with size of $P \times Q$ and $Q \times R$, so that they can be loaded into the cache for faster computations. Constants of $P$, $Q$ and $R$ are determined by the host machine configuration. More details about GEMM can be found in Appendix \ref{app_gemm}. 
% Our testbed (Section \ref{sec_setup}) adopts $P=320$, $Q=320$, $R=104512$.
% As $R$ is generally larger than $n$ in NAS models, we assume $loop_1$ is performed only once. More details about GEMM can be found in Appendix \ref{app_gemm}.

Following Cache Telepathy \cite{yan2020cache}, we take the \emph{itcopy} and \emph{oncopy} APIs in OpenBLAS as the monitoring targets. Since these two APIs are used to load matrix data into the cache, we can analyze the access pattern to them to reveal the dimension information of computing matrix. Besides, the variance of API access pattern also leaks the type of running operation. 
Figure \ref{fig_patterns} illustrates the leakage patterns of four representative operations with a sampling interval of 2000 CPU cycles. Different operations have distinct patterns of side-channel leakage. By observing such patterns, we can identify the type of the operation.

Finally, we derive the hyper-parameters of each operation based on the inferred matrix dimension. The relationships between the hyper-parameters of various operations and the dimensions of the transformed matrices are summarized in Table \ref{table_conv}. We present both the general calculations of the hyper-parameters as well as the ones specifically for NAS models. Below we give detailed descriptions on the recovery of each NAS operation.

% \begin{figure*}[hbt]
%   \centering
%   \includegraphics[width=0.85\linewidth]{./fig/fig_patterns.pdf}
%   \caption{Side channel patterns of operations in NAS.}
%   \label{fig_patterns}
% \end{figure*}

% Second, we utilize the technique in \cite{yan2020cache} to derive the range of the matrix dimension $(m, n, k)$ from $iter_n$, based on the equations: $iter_1 \equiv 1$, $iter_2=\lceil k/Q \rceil$, $iter_3=\lceil (m-P)/P \rceil$ and $iter_4=\lceil n/3UNROLL \rceil$. 
% Note that the final two iterations of each loop are actually assigned with two equal-size blocks, rather than blocks of size $m$ (or $n, k$). This does not make big differences on the derivation.
% Then we deduce the possible values of matrix dimension from the range, based on the constraints of NAS models. 

% Third, we derive the hyper-parameters of each operation based on the matrix dimension. The relationships between the hyper-parameters of various operations and the dimensions of the transformed matrices are summarized in Table \ref{table_conv}. We present both the general calculations of the hyper-parameters as well as the ones specifically for NAS models. Below we give detailed descriptions of the above three-step analysis for each operation.

\subsection{Recovery of NAS Operations} \label{sec_pattern}
%Below we describe how to identify each type of the operation, and recover the corresponding hyper-parameters.

%Besides, different operations can cause distinct patterns of side-channel leakage. By combining the leakage pattern and deduced hyper-parameters, we can identify the critical operations (i.e., watermark) from the target DNN model with high fidelity. Below, we describe how we analyze the patterns of different operations.

%To simplify the watermark verification, we make a comprehensive study about the side-channel patterns of common operations used in numerous 

%After identifying the operations based on the leakage patterns, we can further recover the hyper-parameters of each operation to verify if it is consistent with the watermark stamps. 

\begin{table*}[htp]
\centering
\resizebox{0.9\linewidth}{!}{
\begin{tabular}{|c|c|c|c|c|c|}
\Xhline{1pt}
\rowcolor{gray!40} Operations & Parameters & Value & Operations  & Parameters & Value \\ 
\Xhline{1pt}
\multirow{2}{*}{Fully Connected} & $C_l$: \# of layers & \# of matrix muls & \multirow{2}{*}{Pooling Layer} & \multirow{2}{*}{pool width/height} & \multirow{2}{*}{$\approx \sqrt{\frac{row(out_{i-1})}{row(in_{i})}}$} \\
\cline{2-3} 
 & $C_n$: \# of neurons & $row(\theta)$ &  &  & \\ 
\Xhline{1pt}
\rowcolor{gray!40} Operations & $D_{i+1}$: Number of Filters & $R_i$: Kernel Size & $P_i$: Padding  & Stride & $d$: Dilated Space \\ 
\Xhline{1pt}
Normal Conv & \multirow{2}{*}{$col(F_i)$} & \multirow{2}{*}{$\sqrt{\frac{col(in_i)}{col(out_{i-1})}}$} & \multirow{4}{*}{\makecell{$diff(row(in_{i}), row(out_{i-1}))$ \\ NAS: $R_i-1$ (non-dilated) \\ $R_i'-1$ (dilated), where \\ $R_i' = R_i + d(R_i-1)$}}  & \multirow{4}{*}{\makecell{$\sqrt{\frac{row(out_{i-1})}{row(in_{i})}}$ \\ NAS: = 1 (normal cells) \\ = 2 (reduction cells)}} & 0 \\
\cline{1-1} \cline{6-6}
Dilated Conv &  &  &   & & d \\
\cline{1-3}  \cline{6-6}
\multirow{1}{*}{Separable Conv} & \multirow{2}{*}{\makecell{Filters \circled{1}: \# of same matrix muls \\ Filters \circled{2}: $col(F_i)$}} & \multirow{2}{*}{\makecell{Filters \circled{1}: $\sqrt{row(F_i)}$ \\ Filters \circled{2}: $1$}} &  & & \multirow{1}{*}{0} \\
\cline{1-1}  \cline{6-6}
Dil-Sep Conv &  &  &   & & d \\
\hline
\end{tabular}}
\caption{Mapping between operation hyper-parameters and matrix dimensions. }\label{table_conv}
\vspace{-5pt}
\end{table*}

\noindent\textbf{Fully connected (FC) layer.} 
%This operation is ubiquitous in modern neural networks for analyzing the extracted features. 
This operation can be transformed to the multiplication of a learnable weight matrix $\theta$ ($m \times k$) and an input matrix $in$ ($k \times n$), to generate the output matrix $out$ ($m \times n$). $m$ denotes the number of neurons in the layer; $k$ denotes the size of the input vector; and $n$ reveals the batch size of the input vectors. Hence, with the possible values of $(m, n, k)$ derived from the iteration counts of \emph{itcopy} and \emph{oncopy}, hyper-parameters (e.g., neurons number, input size) of the FC layer can be recovered. The number of FC layers in the model can also be recovered by counting the number of matrix multiplications. Figure \ref{fig_patterns}(a) shows the pattern of a classifier with two FC layers, where two separate clusters can be easily identified.
% the first layer has 1024 neurons and the second layer has 100 neurons.
% From this pattern, it is easy to identify these two FC layers. 
% The first layer takes as input a batch of 12 vectors of size 512, and it has m=1024, n=12, k=512. We can infer the range of $(m,n,k)$ based on the number of iterations in each loop: $iter_2=2$ (i.e., number of pattern $I-O-3I$), $iter_4=1$ and $iter_3=3$.
% % m=1024, n=12, k=512

\noindent\textbf{Normal convolution.}
Although this operation was adopted in earlier NAS methods \cite{real2019regularized, zoph2018learning}, recent works \cite{liu2018darts,dong2019searching,chu2020fair} removed it from the search space as it is hardly used in the searched cells. 
% Hence, the normal convolution is only applied in the preprocessing stage of the NAS model, following the structure of \emph{ReLU-Conv-BN} \cite{liu2018darts}. 
However, this operation is the basis of the following complex convolutions. So it is necessary to perform detailed analysis about it. 

\begin{figure}[t]
  \centering
  \includegraphics[width=0.8\linewidth]{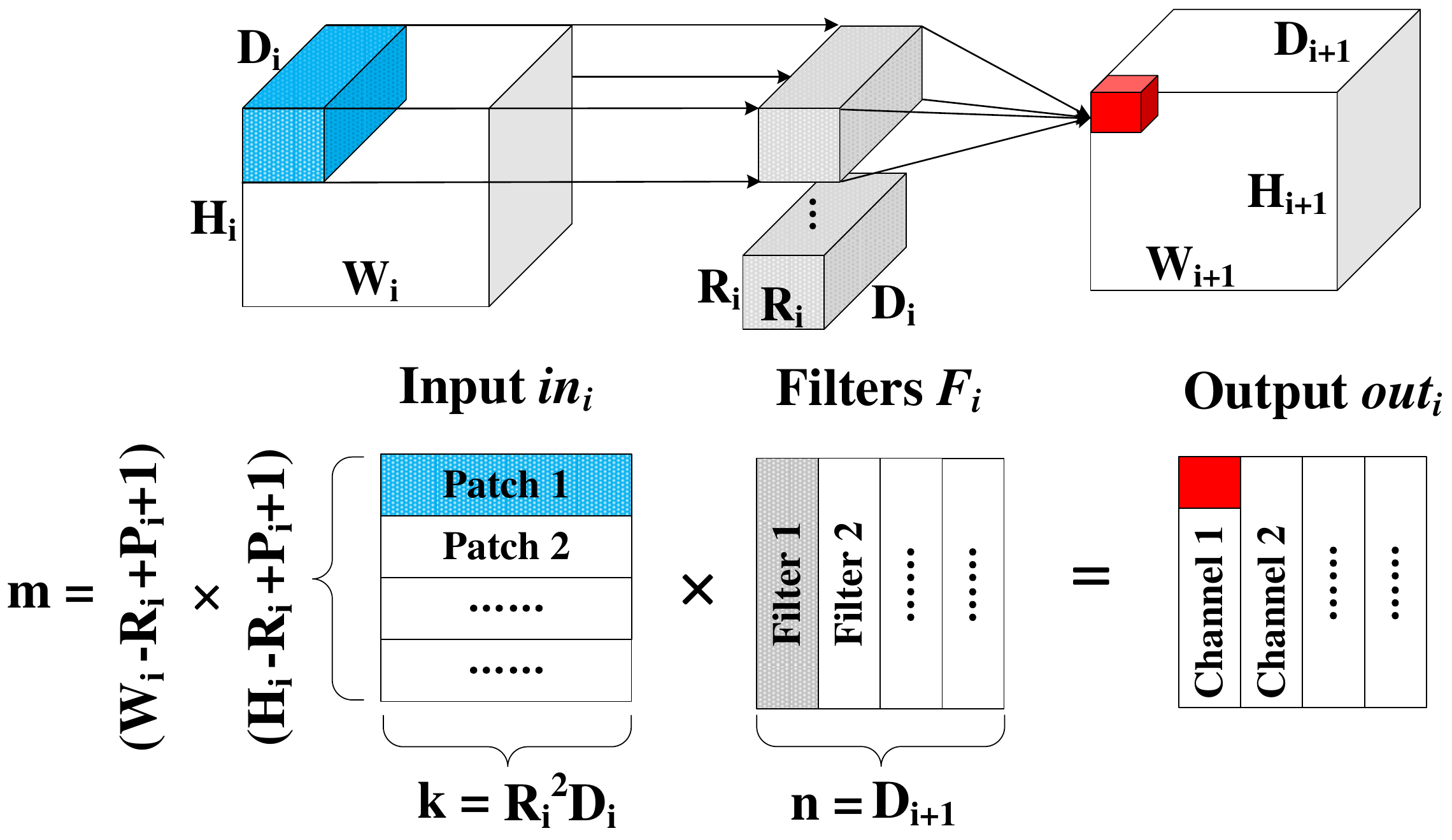}
  \vspace{-5pt}
  \caption{Implementing a convolution operation as matrix multiplication}
  \vspace{-5pt}
  \label{fig_conv}
\end{figure}

Figure \ref{fig_conv} shows the structure of a normal convolution at the $i$-th layer (upper part), and how it is transformed to a matrix multiplication (lower part). 
% Specifically, the input tensor has a dimension of $H_i \times W_i \times D_i$, where $D_i$ is the number of input channels. It also has $D_{i+1}$ filters with the size of $R_i^2 \times D_i$. 
Each patch in the input tensor is stretched as a row of matrix $in_i$, and each filter is stretched as a column of matrix $F_i$. Hence, the number of filters $D_{i+1}$ can be recovered from the column size $n$ of the filter matrix $F_i$. The kernel size $R_i$ can be revealed from the column size $k=R_i^2D_i$ of the matrix $in_i$, as we assume $D_i$ has been obtained from the previous layer. With the recovered $R_i$, the padding size $P_i$ can be inferred as the difference between the row sizes of $out_{i-1}$ and $in_i$, which are $W_i \times H_i$ and $(W_i-R_i+P_i+1)(H_i-R_i+P_i+1)$, respectively. The stride can be deduced based on the modification between the input size and output size of the convolution.
In a NAS model, the convolved feature maps are padded to preserve their spatial resolution, so we have $P_i=R_i-1$. A normal cell takes a stride of 1, while a reduction cell takes a stride of 2. 

% The patch size is equal to the filter size $R_i^2D_i$. There are totally $(W_i-R_i+P_i+1)(H_i-R_i+P_i+1)$ patches in the input volume, where $P_i$ is the padding size. The column number of $F_i$ is the same as the number of filters $D_{i+1}$. Then the multiplication of $in_i$ and $F_i$ generates the output matrix $out_i$. Such order in which $in_i$ and $F_i$ are multiplied is common in many popular DL frameworks, e.g., Pytorch.

% As shown in Figure \ref{fig_conv}, the $i$-th normal convolution layer is transformed to a multiplication of an input matrix $in_i$ and filter matrix $F_i$. Hence,
% the number of filters (i.e., the depth of the input to the next layer $D_{i+1}$) can be first recovered, which is equal to the column size $n$ of the filter matrix $F_i$. Then the kernel size $R_i$ can also be recovered from the row size $k=R_i^2D_i$ of the matrix $F_i$. We assume the input depth $D_i$ is known, as it can be obtained from the previous layer. Based on $R_i$, the padding size $P_i$ can be inferred as the difference between the row size of the output matrix $out_{i-1}$ in the layer $(i-1)$, which is $W_i \times H_i$, and the row size of the input matrix $in_i$, which is $(W_i-R_i+P_i+1)(H_i-R_i+P_i+1)$. In the NAS scenario, 
% the convolved feature maps are padded to preserve their spatial resolution, so we have $P_i=R_i-1$. The stride can be deduced based on the modification between the input size and output size of the convolution. For NAS models a normal cell takes a stride of 1, while a reduction cell takes a  stride of 2. 

In terms of the leakage pattern, a normal convolution is hard to be distinguished from a FC layer, as both of their accesses to the \emph{itcopy} and \emph{oncopy} functions can be denoted as $xI-yO-zI$, where $(x, y, z)$ indicate the repeated times of the functions, determined by the operation hyper-parameters. This is why Cache Telepathy \cite{yan2020cache} needs to know the architecture family of the target DNN to distinguish the operations.
Figure \ref{fig_patterns}(b) shows the leakage pattern of a normal convolution.
% ($H_i=W_i=32$, $D_i=33$, $R_i=3$, $P_i=2$ and $D_{i+1}=132$), which has $iter_2=1$, $iter_4=11$ and $iter_3=3$. 
% Note that the first two red nodes (interval 409 and 410) can be treated as one iteration, as they occur in a very short period and generated by side-channel noise. 
In the NAS scenario, since the normal convolution is generally used at the preprocessing stage, while the FC layer is adopted as the classifier at the end, they can be distinguished based on their locations.

% Similar with the FC layer, matrix dimensions transformed from the convolutions can be derived from $iter_n$ in Algorithm \ref{alg_gemm}.
% Figure \ref{fig_patterns}(b) shows the leakage pattern of a normal convolution, with $H_i=W_i=32$, $D_i=33$, $R_i=3$, $P_i=2$ and $D_{i+1}=132$.
% The operation has $iter_2=1$, $iter_4=11$ and $iter_3=3$, which helps us to deduce the range of matrix dimension, while the actual values are m=1024, n=132, k=297. Particularly, the first two red nodes at around the 400th interval can be treated as one iteration, as they are monitored within a very short period and generated by side-channel noise. 

% In terms of the leakage pattern, a normal convolution is hard to be distinguished from a FC layer, as both of their accesses to the \emph{itcopy} and \emph{oncopy} functions can be denoted as $xI-yO-zI$, where $(x, y, z)$ indicate the repeated times of the functions, determined by the operation hyper-parameters. This is why Cache Telepathy \cite{yan2020cache} needs to know the architecture family of the target DNN to distinguish the operations. In our scenario, since the normal convolution is generally used at the pre-processing stage, while the FC layer is adopted as the classifier at the end, they can be distinguished based on their locations. 

\noindent\textbf{Dilated convolution.}
This operation is a variant of the normal convolution, which inflates the kernel by inserting spaces between each kernel element. We use the dilated space $d$ to denote the number of spaces inserted between two adjacent elements in the kernel.
%where $d$ is 0 for a normal convolution. 
The conversion from the hyper-parameters of a dilated convolution to the matrix dimension is similar with the normal convolution. The only difference is the row size $m$ of the input matrix $in_i$, i.e., the number of patches. Due to the inserted spaces in the kernel, although the kernel size is still $R_i^2$, the actual size covered by the dilated kernel becomes $R_i'^2$, where $R_i' = R_i + d(R_i-1)$. This changes the number of patches to $(W_i-R_i'+P_i+1)(H_i-R_i'+P_i+1)$. As a dilated convolution is normally implemented as a dilated separable convolution in practical NAS methods \cite{liu2018darts,dong2019searching}, the leakage pattern of the operation will be discussed with the dilated separable convolution. 

\noindent\textbf{Separable convolution.}
According to \cite{yan2020cache}, the number of consecutive matrix multiplications with the same pattern reveals the batch number of a normal convolution. However, we find this does not hold in the separable convolution, or precisely, the depth-wise separable convolution used in NAS. This is because the separable convolution decomposes a convolution into multiple separate operations, which can incur the same conclusion that the number of the same patterns equals to the number of input channels. 

\begin{figure}[t]
  \centering
  \includegraphics[width=0.8\linewidth]{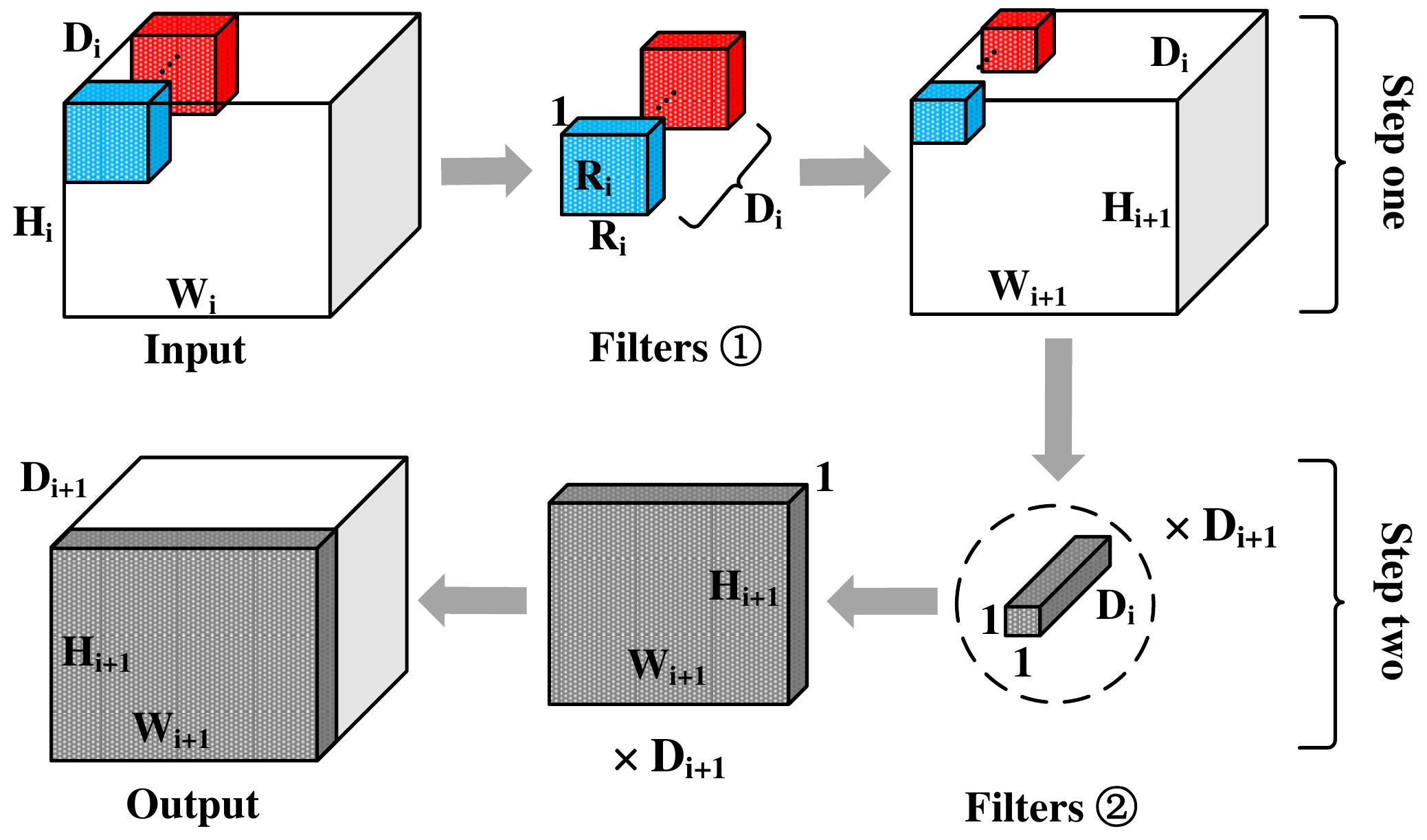}
  \vspace{-5pt}
  \caption{Procedure of separable convolutions.}
  \label{fig_sepconv}
  \vspace{-5pt}
\end{figure}

A separable convolution aims to achieve more efficient computation with less complexity by separating the filters. 
% There are two main types: (1) spatial separable convolution is rarely used in DNN models as it needs to divide the kernel into two smaller ones; (2) depth-wise separable convolution is more commonly adopted in deep learning, especially the NAS model. For simplicity, we focus on the depth-wise separable convolution, and ignore ``depth-wise'' in this paper. 
Figure \ref{fig_sepconv} shows a two-step procedure of a separable convolution. 
The first step uses $D_i$ filters (Filters \circled{1}) to transform the input to an intermediate tensor, where each filter only convolves one input channel to generate one output channel. It can be regarded as $D_i$ normal convolutions, with the input channel size of 1 and the filter size of $R_i^2\times1$. These computations are further transformed to $D_i$ consecutive matrix multiplications with the same pattern, which is similar as a normal convolution with the batch size of $D_i$. 
% The first step uses $D_i$ filters (Filters \circled{1}) of size $R_i^2\times1$ to transform the input to an intermediate tensor of size $W_{i+1} \times H_{i+1} \times D_i$, where each filter only convolves one input channel to generate one output channel. Each of these normal convolutions has $D_i=1$, indicating the dimensions of the transformed matrices: the row size $m$ of the input matrix $in_i$ is still $(W_i-R_i+P_i+1)(H_i-R_i+P_i+1)$; the column size $k$ changes to $R_i^2$; the column size $n$ of the filter matrix $F_i$ is 1. As a result, the $D_i$ normal convolutions are transformed to $D_i$ consecutive matrix multiplications with the same pattern, which is similar as a normal convolution with the batch size of $D_i$. 
But the separable convolution has much shorter intervals between two matrix multiplications, as they are parts of the whole convolution, rather than independent operations. In the second step, a normal convolution with $D_{i+1}$ filters (Filters \circled{2}) of size $1^2 \times D_i$ is applied to the intermediate tensor to generate the final output. 
% where each filter provides an output channel. The kernel size of these $D_{i+1}$ filters is $R_i=1$. So we have $m=(W_i+P_i)(H_i+P_i)$, $k=D_i$ and $n=D_{i+1}$. 

%Compared with the normal convolution in Figure \ref{fig_conv}, which requires $D_{i+1}$ filters of size $R_i^2 \times D_i$, the separable convolution needs much less computation.

In summary, the leakage pattern of the separable convolution is fairly distinguishable, which contains $D_i$ consecutive clusters and one individual cluster at the end. Note that in a NAS model, the separable convolution is always applied twice \cite{zoph2018learning, real2019regularized, liu2018progressive, liu2018darts, dong2019searching} to improve the performance, which makes its leakage pattern more recognizable. Figure \ref{fig_patterns}(c) shows the trace of a separable convolution.
% that takes 12-channel $32 \times 32$ tensor and uses 12 filters with the size of $3 \times 3 \times 1$ (Filters \textcircled{1}) and 33 $1 \times 1 \times 12$ filters (Filters \textcircled{2}). 
There are clearly two parts following the same pattern, corresponding to the two occurrences of the operation. Each part contains 12 consecutive same-pattern clusters to reveal $D_i=12$, and an individual cluster denoting the last $1 \times 1$ convolution. 

\noindent\textbf{Dilated separable (DS) convolution.}
This operation is the practical implementation of a dilated convolution in NAS. The DS convolution only introduces a new variable, the dilated space $d$, from the separable convolution. Hence, this operation has similar matrix transformation and leakage pattern as the separable convolution, except for two differences. 
First, $R_i$ is changed to $R_i' = R_i + d(R_i-1)$ in calculating the number of patches $m=(W_i-R_i+P_i+1)(H_i-R_i+P_i+1)$ in Step One.  
% First, $R_i$ in the row size $m=(W_i-R_i+P_i+1)(H_i-R_i+P_i+1)$ of the input matrix $in_i$ in Step One changes to $R_i' = R_i + d(R_i-1)$. 
Second, a DS convolution needs much shorter execution time. Figure \ref{fig_patterns}(d) shows the leakage pattern of a DS convolution with the same hyper-parameters as a separable convolution depicted in Figure \ref{fig_patterns}(c), except that the dilated space $d=1$. It is easy to see the performance advantage of the DS convolution (8400 intervals) over the separable convolution (10000 intervals) under the same configurations. The reason is that the input matrix in a DS convolution contains more padding zeroes to reduce the computation complexity. Besides, the DS convolution does not need to be performed twice, which also helps us distinguish it from a separable one. 

\noindent\textbf{Skip connect.} The operation is also called \emph{identity} in the NAS search space, which just sends $out_{i}$ to $in_{j}$ without any processing. This operation cannot be directly detected from the side-channel leakage trace, as it does not invoke any GEMM computations. While \cite{yan2020cache} argues the skip can be identified as it causes a longer latency due to the introduction of an extra merge operation, it is not feasible in a NAS model. This is because in a cell, each node has an add operation of two inputs and the skip operation does not invoke any extra operations. So there is no obvious difference between the latency of skip and the normal inter-GEMM intervals. 
%As a result, we cannot reliably identify the skip operation based on the analysis of execution latency. 
% Our experiments show that while the skip connect cannot be distinguished in a CNN model, it can still be identified in an RNN model. More details can be found in Section \ref{sec_eva}. 

\noindent\textbf{Pooling.} 
%This layer is normally used to extract global features and provide local invariance. In this paper, 
We assume the width and height of the pooling operation is the same, which is default in all practical implementations. Given that pooling can reduce the size of the input matrix $in_i$ from the last output matrix $out_{i-1}$, the size of the pooling layer can be obtained by performing square root over the quotient of the number of rows in $out_{i-1}$ and $in_i$. In general, pooling and non-unit striding cannot be distinguished as they both reduce the matrix size. However, in a NAS model, non-unit striding is only used in reduction cells which can double the channels. This information can be used for identification. Pooling cannot be directly detected as it does not invoke any matrix multiplications in GEMM. But it can introduce much longer latency (nearly 1.5$\times$ of the normal inter-GEMM latency) for other computations. Hence, we can identify this operation by monitoring the matrix size and execution intervals. While monitoring the BLAS library can only tell the existence of the pooling operation, the type can be revealed by monitoring the corresponding pooling functions in the deep learning framework. 

\noindent\textbf{Other DNN components.}
Besides the above operations, other common components like batch normalization, dropout and activation functions are also critical to the model performance. Our method can be generalized to watermark these components as well, by just changing the monitored library targets. For instance, to protect activation functions, e.g., \emph{relu} and \emph{tanh}, we can monitor accesses to the corresponding function APIs in Pytorch. In Appendix \ref{app_monitor}, we give more details about the monitored code lines in Table \ref{table_rnn} and an example side-channel trace of monitored activation functions in Figure \ref{fig_rnn}.
\section{Evaluation} \label{sec_eva}
\subsection{Experimental Setup} \label{sec_setup}
\textbf{Testbed.} 
Our approach is general for different deep learning frameworks and libraries. Without loss of generality, we adopt Pytorch (1.7.0) and OpenBLAS (0.3.13), deployed in Ubuntu 18.04 with a kernel version of 4.15.0. Evaluations are performed on a workstation of Dell Precision T5810 (6-core Intel Xeon E5 processor, 32GB DDR4 memory). The processor has core-private 32KB L1 caches, 256KB L2 caches and a shared 15MB last level cache. 

\noindent\textbf{NAS implementation.} 
Our scheme is independent of the search strategy, and can be applied to all cell-based NAS methods. We mainly focus on the CNN tasks, and select a state-of-the-art NAS method GDAS \cite{dong2019searching}, which can produce qualified network designs within five GPU hours. We follow the default configurations to perform NAS \cite{zoph2018learning,dong2019searching}: the search space of a CNN cell contains: \textit{identity (skip), 3$\times$3 and 5$\times$5 separable convolutions (SC), 3$\times$3 and 5$\times$5 dilated separable convolutions (DS), 3$\times$3 average pooling (AP), 3$\times$3 max pooling (MP)}. The discovered cells are then stacked to construct DNN models. We adopt CIFAR10 as the proxy dataset to search the architecture, and train CNN models over different datasets, e.g., CIFAR10, CIFAR100, ImageNet. Technical details about cell search and model training can be found in Appendix \ref{app_nas}.
% We also consider watermarking RNN models (Section \ref{sec:watermark-rnn}). We choose the DARTS \cite{liu2018darts} method, which can generate models within six GPU hours. The search space of an RNN cell contains the operations of \emph{tanh}, \emph{relu}, \emph{sigmoid} activations, identity and zeroize. We use the PTB dataset to search and train models. 

\noindent\textbf{Side channel extraction.}
To capture the side-channel leakage of CNN models, we monitor the \emph{itcopy} and \emph{oncopy} functions in OpenBLAS. 
% For RNN models, we monitor the activation functions (\emph{tanh}, \emph{relu} and \emph{sigmoid}) in Pytorch, since executions in OpenBLAS do not leak information about the models. 
We adopt the \textsc{Flush+Reload} side-channel technique \cite{yarom2014flush+}, but other methods can achieve our goal as well. We inspect the cache lines storing these functions at a granularity of 2000 CPU cycles to obtain accurate information. Details about the monitored code locations can be found in Table \ref{table_rnn} in Appendix \ref{app_monitor}.

\subsection{Effectiveness} \label{sec_eff}
% \shangwei{this subsection is too long}

\begin{table}[t]
\centering
\resizebox{0.9\linewidth}{!}{
\begin{tabular}{|c|c|c|c|c|c|}
\hline
\multirow{2}{*}{$mk_n$} & $s_{en}$ & $c_{i-2} \to \mathcal{N}_0$   & $\mathcal{N}_0 \to \mathcal{N}_1$   &  $\mathcal{N}_1 \to \mathcal{N}_2$  & $\mathcal{N}_2 \to \mathcal{N}_3$  \\ \cline{2-6} 
                  & $s_{on}$  &  $3 \times 3$ AP   &  $5 \times 5$ SC  &  $3 \times 3$ DS  &   $3 \times 3$ SC  \\ \hline
\multirow{2}{*}{$mk_r$} & $s_{er}$ &   $c_{i-1} \to \mathcal{N}_0$   & $\mathcal{N}_0 \to \mathcal{N}_1$   &  $\mathcal{N}_1 \to \mathcal{N}_2$  & $\mathcal{N}_2 \to \mathcal{N}_3$   \\ \cline{2-6} 
                  & $s_{or}$ &  $3 \times 3$ DS   &  $3 \times 3$ SC   &  $3 \times 3$ SC   &  skip   \\ \hline
\end{tabular}}
\caption{An example of the marking key $mk$.} \label{table_effkey}
\vspace{-10pt}
\end{table}

\subsubsection{Key Generation}
A NAS method generally considers two types of cells. So we set the same stamp for each type. Then the marking key can be denoted as $mk=(mk_n, mk_r)$, where $mk_n=\{s_{en}:s_{on}\}$ and $mk_r=\{s_{er}:s_{or}\}$ represent the stamps embedded to the normal and reduction cells, respectively. Each cell has four computation nodes ($\mathcal{B}=4$), and we set the number of stamp edges $n_s=4$ for both cells, indicating four causal edges in each cell are fixed and attached with random operations. We follow Algorithm \ref{alg1} to generate one example of $mk$ (Table \ref{table_effkey}). The verification key $vk=(vk_n, vk_r)$ is also recorded, where $vk_n=s_{on}$ and $vk_r=s_{or}$. 

\begin{figure*}[t]
  \centering
  \begin{subfigure}{0.4\linewidth}
    \centering
    \includegraphics[width=\linewidth]{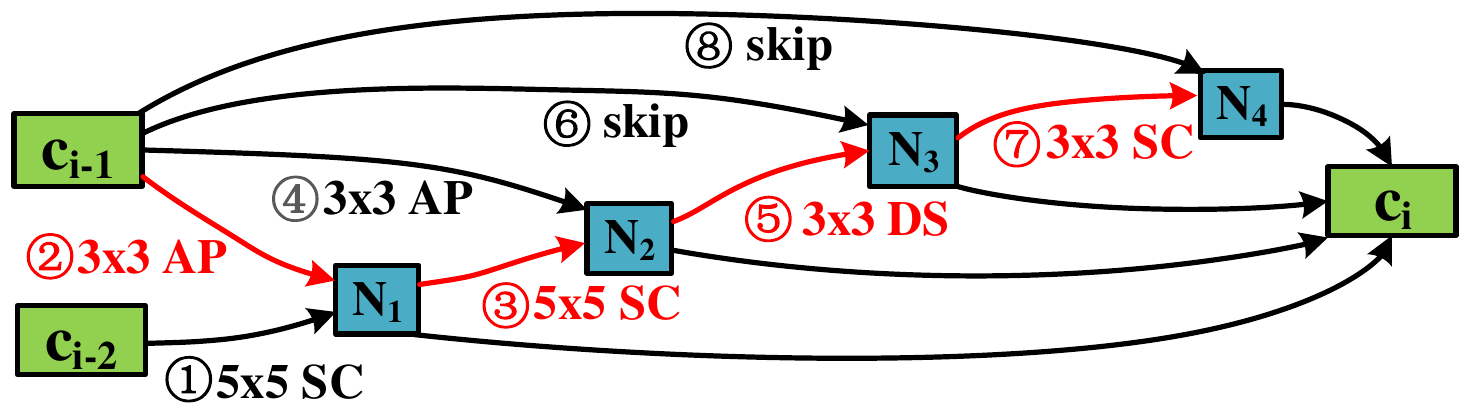}
    \vspace{-10pt}
    \caption{Normal cell}
    \label{fig_norcell}
  \end{subfigure}%
%   \hfill
  \begin{subfigure}{0.4\linewidth}
    \centering
    \includegraphics[width=\linewidth]{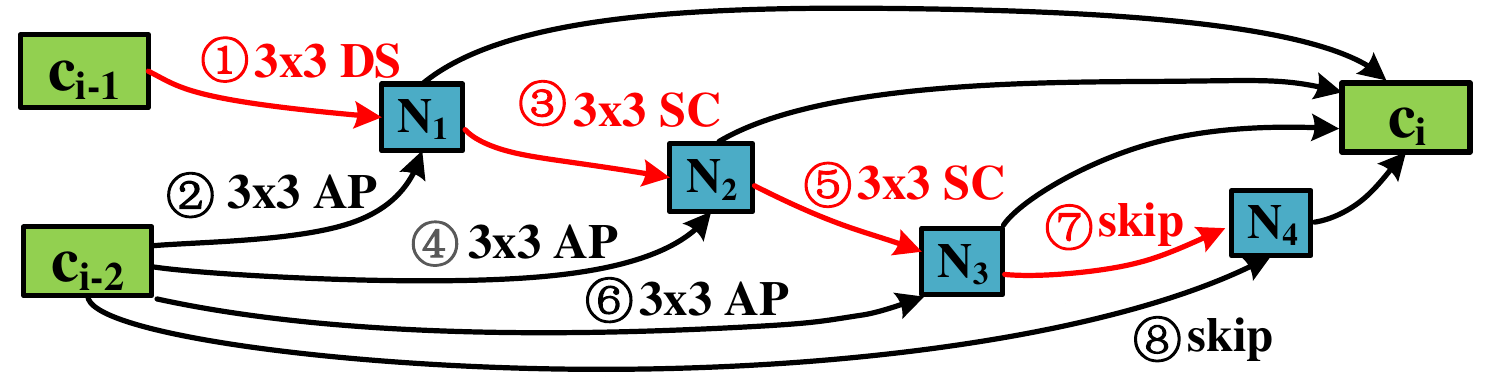}
    \vspace{-10pt}
    \caption{Reduction cell}
    \label{fig_reducell}
  \end{subfigure}
    %   \vspace{-10pt}
  \caption{Architectures of the searched cells. $c_{i-1}$ and $c_{i-2}$ are the inputs from the previous cells.}
      \vspace{-5pt}
  \label{fig_cells}
\end{figure*}

\subsubsection{Watermark Embedding}
We follow Algorithm \ref{alg2} to embed the watermark determined by $mk$ to the DNN architecture during the search process. 
Figure \ref{fig_cells} shows the architectures of two cells searched by GDAS, where stamps are marked as red edges, and the computing order of each operation is annotated with numbers. These two cells are further stacked to construct a complete DNN architecture, including three normal blocks (each contains six normal cells) connected by two reduction cells. The pre-processing layer is a normal convolution that extends the number of channels from 3 to 33.
The number of filters is doubled in the reduction cells, and the channel sizes (i.e., filter number) of three normal blocks are set as 33, 66 and 132.  We train the searched architecture over CIFAR10 for 300 epochs to achieve a 3.52\% error rate on the validation dataset. This is just slightly higher than the baseline (3.32\%), where all connections participate in the search process. This shows the usability of our watermarking scheme.

\begin{figure}[t]
  \centering
  \includegraphics[width=\linewidth]{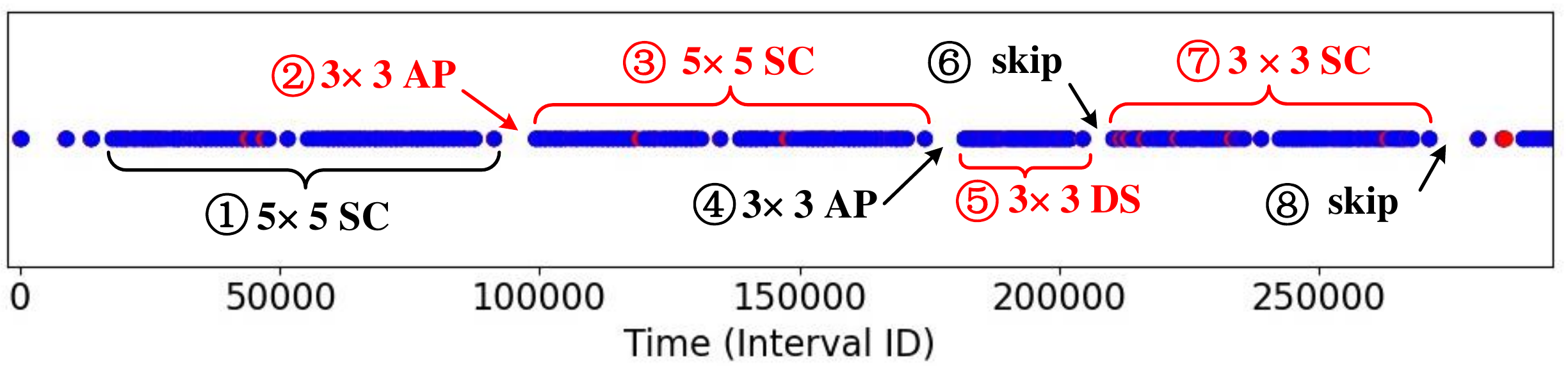}
%   \vspace{-20pt}
  \caption{A side-channel trace of the first normal cell.}
  \label{fig_trace}
  \vspace{-5pt}
\end{figure}

\subsubsection{Watermark Extraction and Verification}
Given a suspicious model, we launch a spy process to monitor the activities in OpenBLAS during inference, and collect the side-channel trace. We conduct the following steps to analyze this trace. 

First, we check whether the pattern of the whole trace matches the macro-architecture of a NAS model, i.e., the trace has three blocks, each of which contains six similar leakage windows, and divided by two different leakage windows. 
% An example of a leakage trace can be found in Appendix \ref{app_trace}. 

\begin{figure}[t]
    \centering
  \begin{subfigure}{0.5\linewidth}
    \centering
    \includegraphics[width=\linewidth]{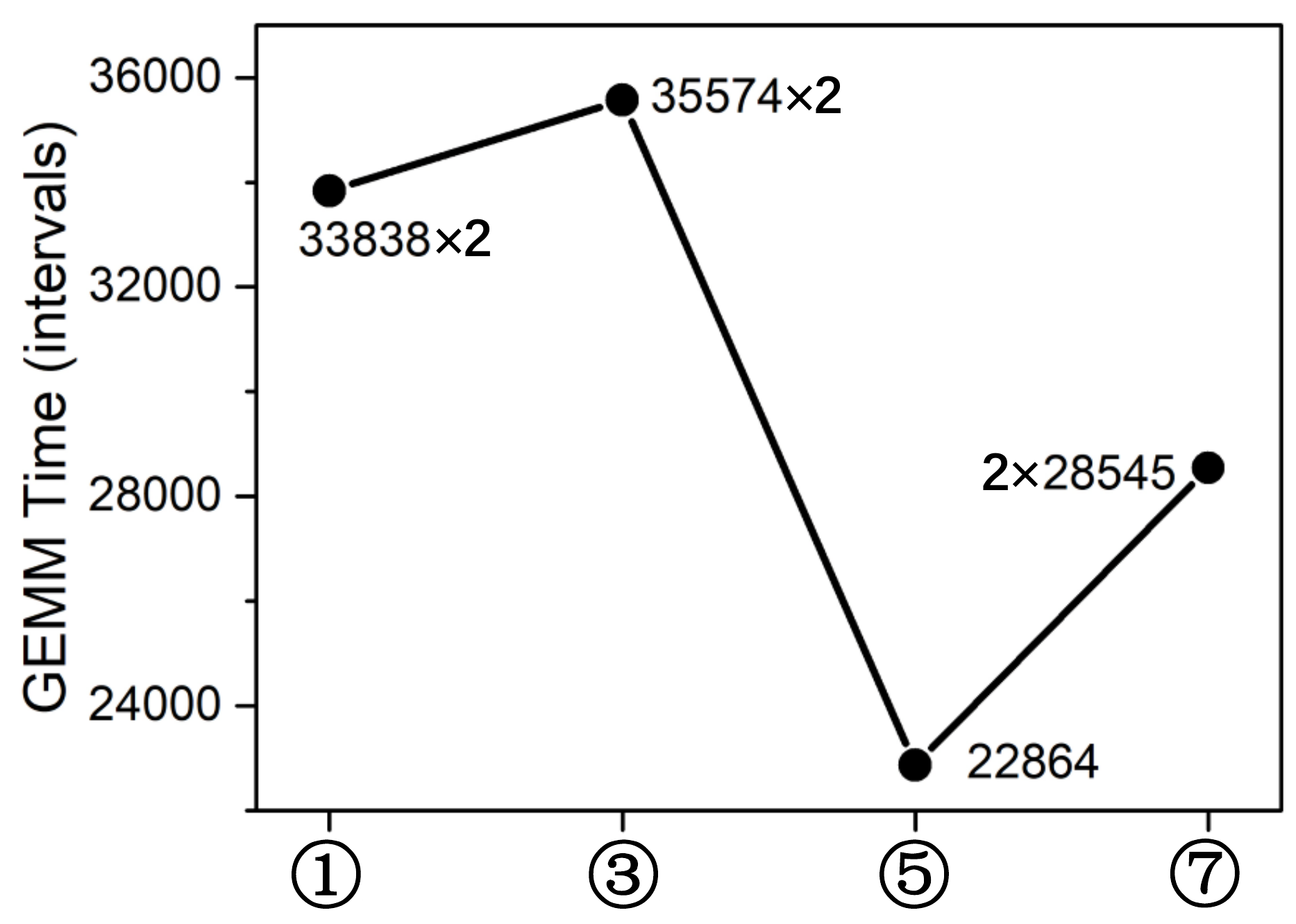}
      \vspace{-5pt}
    \caption{GEMM operations}
    \label{fig_gemmtime}
  \end{subfigure}%
%   \hfill
  \begin{subfigure}{0.5\linewidth}
    \centering
    \includegraphics[width=\linewidth]{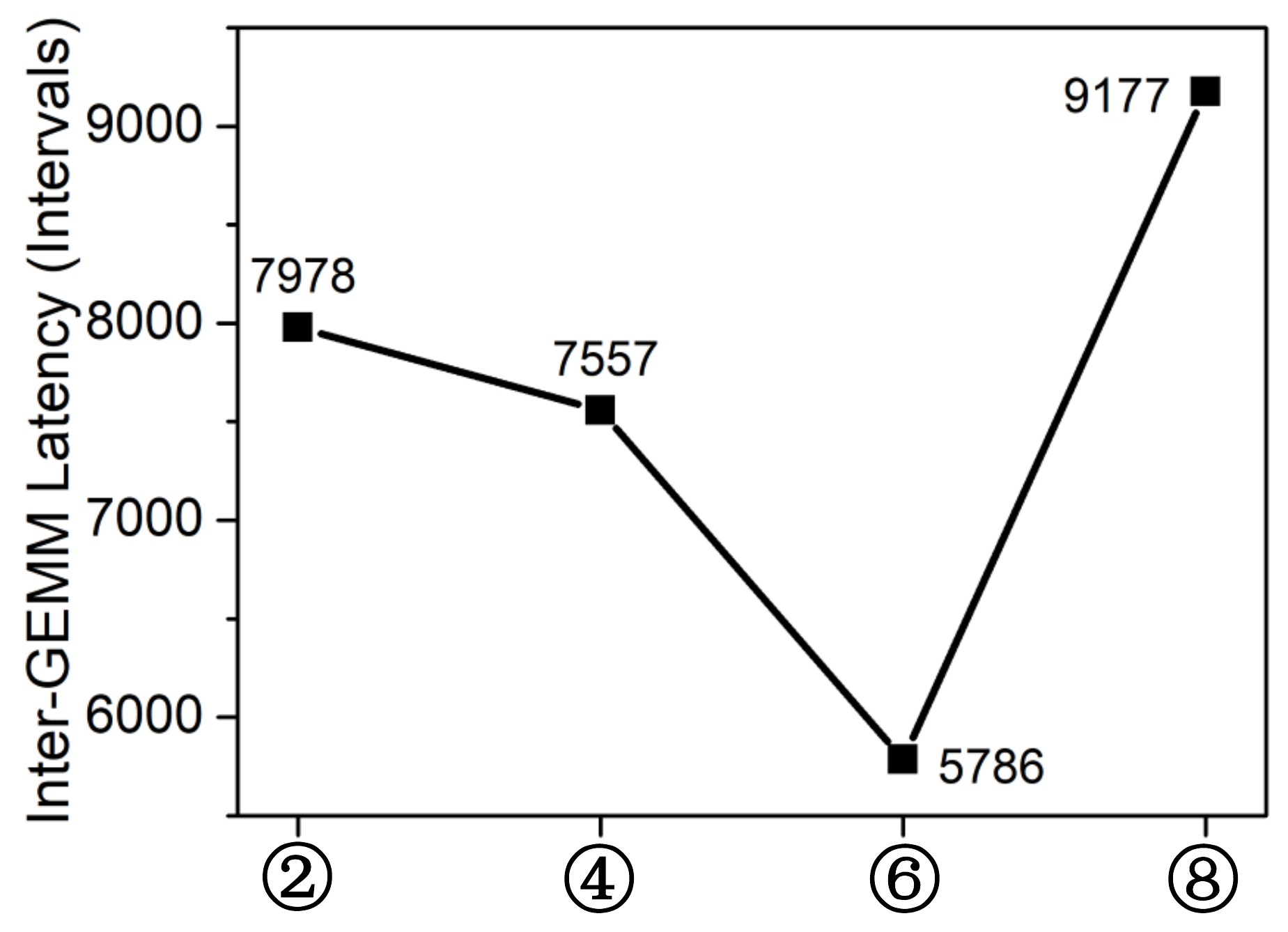}
      \vspace{-5pt}
    \caption{Inter-GEMM latency}
    \label{fig_interval}
  \end{subfigure}
%   \vspace{-10pt}
  \caption{Execution time of the operations in a cell}
  \label{fig_time}
  \vspace{-5pt}
  \end{figure}

Second, we focus on the internal structure of each cell and check if it contains the fixed operation sequence given by $vk$. Here we only demonstrate the pattern of the first leakage window (i.e., the first normal cell) as an example (Figure \ref{fig_trace}). Other cells can be analyzed in the same way. Recall that in the figure the blue node denotes an access to itcopy and red node denotes an access to oncopy. From this figure, we can observe four large clusters, which can be easily identified according to their leakage patterns that \circled{1}, \circled{3} and \circled{7} are SCs while \circled{5} is a DS.  Figure \ref{fig_gemmtime} shows the measured execution time of these four GEMM operations. An interesting observation is that $5\times5$ convolution takes much longer time than $3\times3$ convolution, because it computes on a larger matrix. Such timing difference enables us to identify the kernel size when the search space is limited. Besides, we can also infer that the channel size is 33, since each operation contains $C=33$ consecutive sub-clusters\footnote{The value of $C$ can be identified if we zoom in Figure \ref{fig_trace}, which is not shown in this paper due to page limit.}. Figure \ref{fig_interval} gives the inter-GEMM latency in the cell. The latency of \circled{2} and \circled{4} is much larger, indicating they are pooling operations. Particularly, the latency of \circled{8} contains two parts: \emph{skip} and interval between two cells. The three small clusters at the beginning of the trace are identified as three normal convolutions used for preprocessing the input. Finally, after identifying the fixed operation sequences ($s_o$) in all cells, we can claim the architecture ownership of the DNN model.  

\begin{figure}[t]
  \centering
  \includegraphics[width=0.75\linewidth]{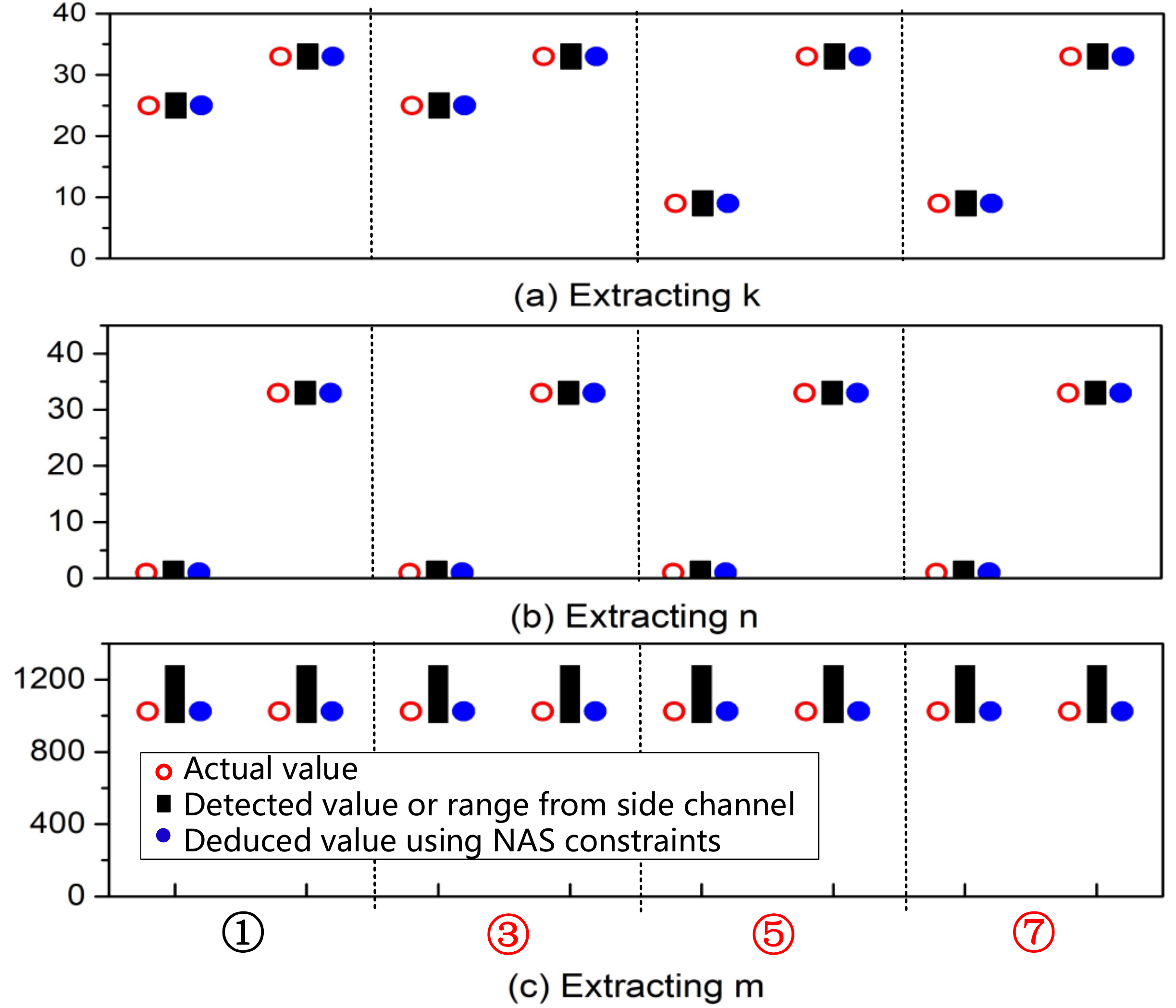}
    % \vspace{-10pt}
  \caption{Extracted values of the matrix parameters $(m, n, k)$.}
  \label{fig_mnk}
  \vspace{-5pt}
\end{figure}

The above analysis can already give us fair verification results. To be more confident, we further recover the remaining hyper-parameters (in particular, the kernel size) based on their matrix dimensions $(m, n, k)$, according to Table \ref{table_conv}. Figure \ref{fig_mnk} shows the values of $(m, n, k)$ extracted from $iter_n$ in the normal cell, where each operation contains two types of normal convolutions. For certain matrix dimensions that cannot be extracted precisely, we empirically deduce their values based on the constraints of NAS models. For instance, $m$ is detected to be between [961, 1280]. We can fix it as $m=1024$ since it denotes the size of input to the cell and $32\times32$ is the most common setting. The value of $n$ can be easily deduced as it equals the channel size. Deduction of $k$ is more difficult, since the filter size $k$ in a NAS model is normally smaller than the GEMM constant in OpenBLAS, it does not leak useful messages in the side-trace trace. However, an interesting observation is that $5\times5$ convolution takes much longer time than $3\times3$ convolution, because it computes on a larger matrix. Such timing difference enables us to identify the kernel size $R_i$ when the search space is limited. Analysis on the reduction cells is similar. 

% \begin{figure}[t]
%   \begin{minipage}[b]{0.49\linewidth}
%     \centering
%   \includegraphics[width=\linewidth]{./fig/fig_trace.pdf}
%     %\vspace{-10pt}
%   \caption{A side-channel trace of the first normal cell.}
%   \label{fig_trace}
%     %\vspace{-10pt}
%   \end{minipage}
%   \hfill
%   \begin{minipage}[b]{0.5\linewidth}
%   \centering
%   \begin{subfigure}{0.5\linewidth}
%     \centering
%     \includegraphics[width=\linewidth]{./fig/fig_gemmtime.pdf}
%     \caption{GEMM operations}
%     \label{fig_gemmtime}
%   \end{subfigure}%
%   \hfill
%   \begin{subfigure}{0.5\linewidth}
%     \centering
%     \includegraphics[width=\linewidth]{./fig/fig_interval.pdf}
%     \caption{Inter-GEMM latency}
%     \label{fig_interval}
%   \end{subfigure}
%     %\vspace{-10pt}
%   \caption{Execution time of the operations}
%   \label{fig_time}
%   \vspace{-5pt}
%   \end{minipage}
%   \end{figure}

% \begin{figure}[t]
%   \centering
%   \begin{subfigure}{0.3\linewidth}
%     \centering
%     \includegraphics[width=\linewidth]{./fig/fig_gemmtime.pdf}
%     \caption{GEMM operations}
%     \label{fig_gemmtime}
%   \end{subfigure}%
%   \hspace*{1pt}
%   \begin{subfigure}{0.3\linewidth}
%     \centering
%     \includegraphics[width=\linewidth]{./fig/fig_interval.pdf}
%     \caption{Inter-GEMM latency}
%     \label{fig_interval}
%   \end{subfigure}
%     %\vspace{-10pt}
%   \caption{Execution time of the operations in a cell}
%   \label{fig_heat}
%   \vspace{-5pt}
% \end{figure}

\begin{figure}[t]
%   \vspace{-5pt}
  \centering
  \begin{subfigure}{0.5\linewidth}
    \centering
    \includegraphics[width=\linewidth]{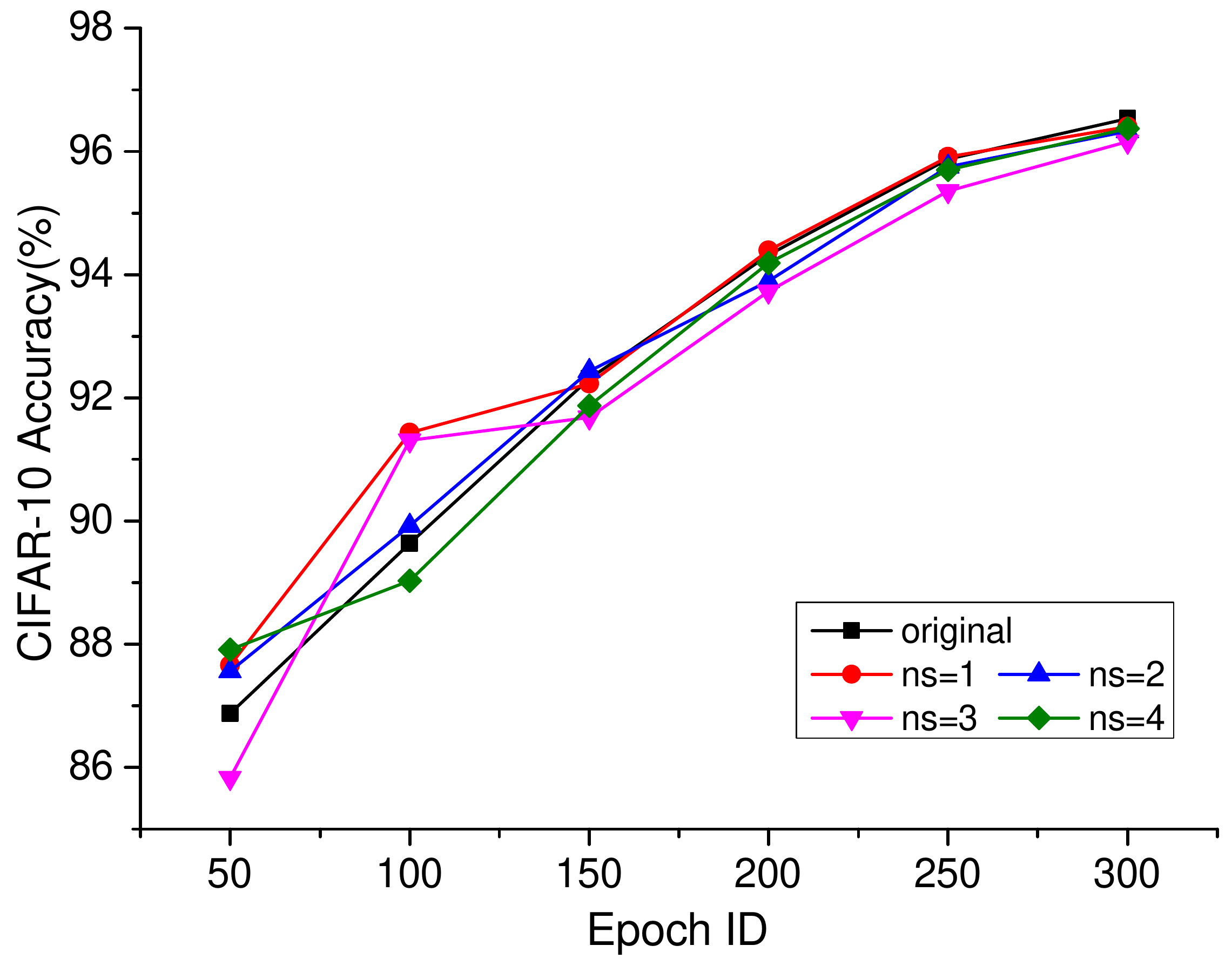}
      \vspace{-10pt}
    \caption{CIFAR-10}
    \label{fig_cifar10}
  \end{subfigure}%
  \hfill
  \begin{subfigure}{0.5\linewidth}
    \centering
    \includegraphics[width=\linewidth]{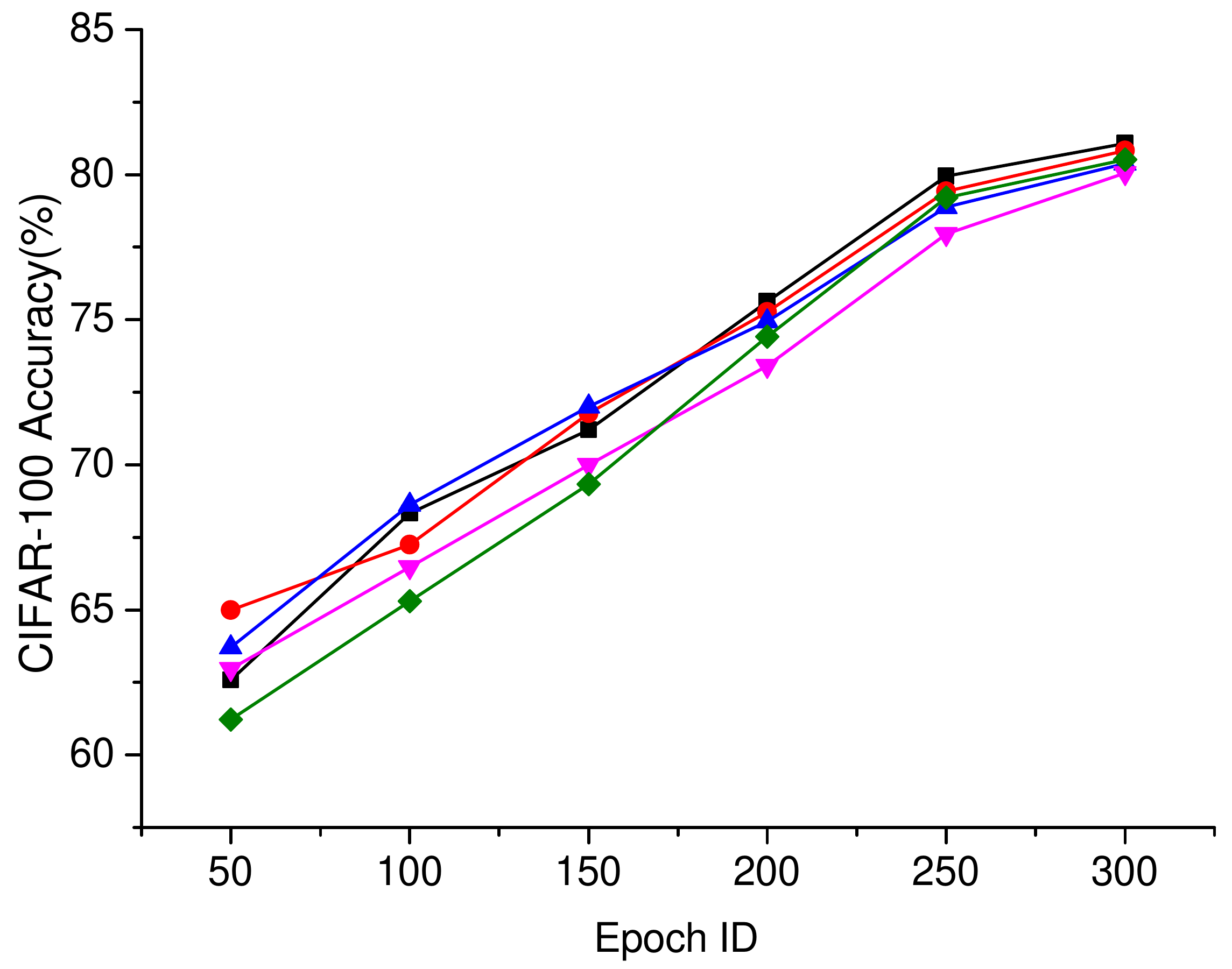}
      \vspace{-10pt}
    \caption{ CIFAR-100}
    \label{fig_cifar100}
  \end{subfigure}
  \vspace{-5pt}
  \caption{Top-1 validation accuracy}
  \label{fig_cifar}
  \vspace{-5pt}
  \end{figure}

\subsection{Usability} \label{sec_funcpresver}
%This property requires the watermarked model has competitive performance with the original one. 
To evaluate the usability property, we vary the number of stamp edges $n_s$ from 1 to 4 to search watermarked architectures. Then we train the models over CIFAR10, CIFAR100 and ImageNet, and measure the validation accuracy. Figure \ref{fig_cifar} shows the average results on CIFAR dataset of five experiments versus the training epochs. 

We observe that models with different stamp sizes have quite distinct performance at epoch 100. Then they gradually converge along with the training process, and finally reach a similar accuracy at epoch 300. For CIFAR10, the accuracy of the original model is 96.53\%, while the watermarked model with the worst performance ($n_s=3$) gives an accuracy of  96.16\%. Similarly for CIFAR100, the baseline accuracy and worst accuracy ($n_s=4$) are 81.07\% and 80.35\%.  We also check this property on ImageNet. Since training an ImageNet model is quite time-consuming (about 12 GPU days), we only measure the accuracies of the original model and two watermarked models ($n_s=2$ and $4$), which are also roughly the same (73.97\%, 73.16\% and 72.73\%). This confirms our watermarking scheme does not affect the usability of the model.

\noindent\textbf{Selection of the stamp size.}
The setting of the stamp size is a trade-off between model usability and watermark reliability. Since our watermark scheme requires the stamp edges to form a dependent path to be robust against operation shuffling attacks (Section \ref{sec_robust}), the largest number of stamps is restricted by the number of nodes in the NAS cell. For conventional NAS architectures, the range of stamp size is [0, 4]. Our evaluation results (Figure \ref{fig_cifar}) indicates that 4 stamp edges incur negligible performance degradation. Therefore, we recommend to adopt this setting in our watermark scheme.

\subsection{Robustness} \label{sec_robust}
We consider the robustness of our watermarking scheme against four types of scenarios.

\noindent\textbf{System noise.}
It is worth noting that the noise in the side-channel traces (e.g., from the system activities, interference with other applications) could possibly make it difficult for the model owner to identify the watermarks. To evaluate this, we follow previous works \cite{hu2020deepsniffer} to inject up to 30\% scales of Gaussian noise into the time interval between events in the side-channel trace, which can well simulate the system noise. We find that it is still feasible to extract the watermarking operations with high fidelity. We conclude that the impact of system noise on operation extraction is actually negligible. The reason behind is that the most important operation features, such as the operation class, channel size and kernel size, are all revealed by analyzing the holistic pattern of side-channel leakage traces. System noise that just disturbs local patterns will not mislead the inference of these operation features. The recovery of matrix dimensions $(m,n,k)$ is indeed affected by side-channel noise, but as we only need to deduce a range of these parameters, such impact is acceptable.
Besides, according to our threat model in Section \ref{sec:threat-model}, the model owner takes control of the host TEE platform, so he can disable other applications on the same machine to further improve the verification reliability.

\begin{figure}[t]
  \centering
  \includegraphics[width=0.9\linewidth]{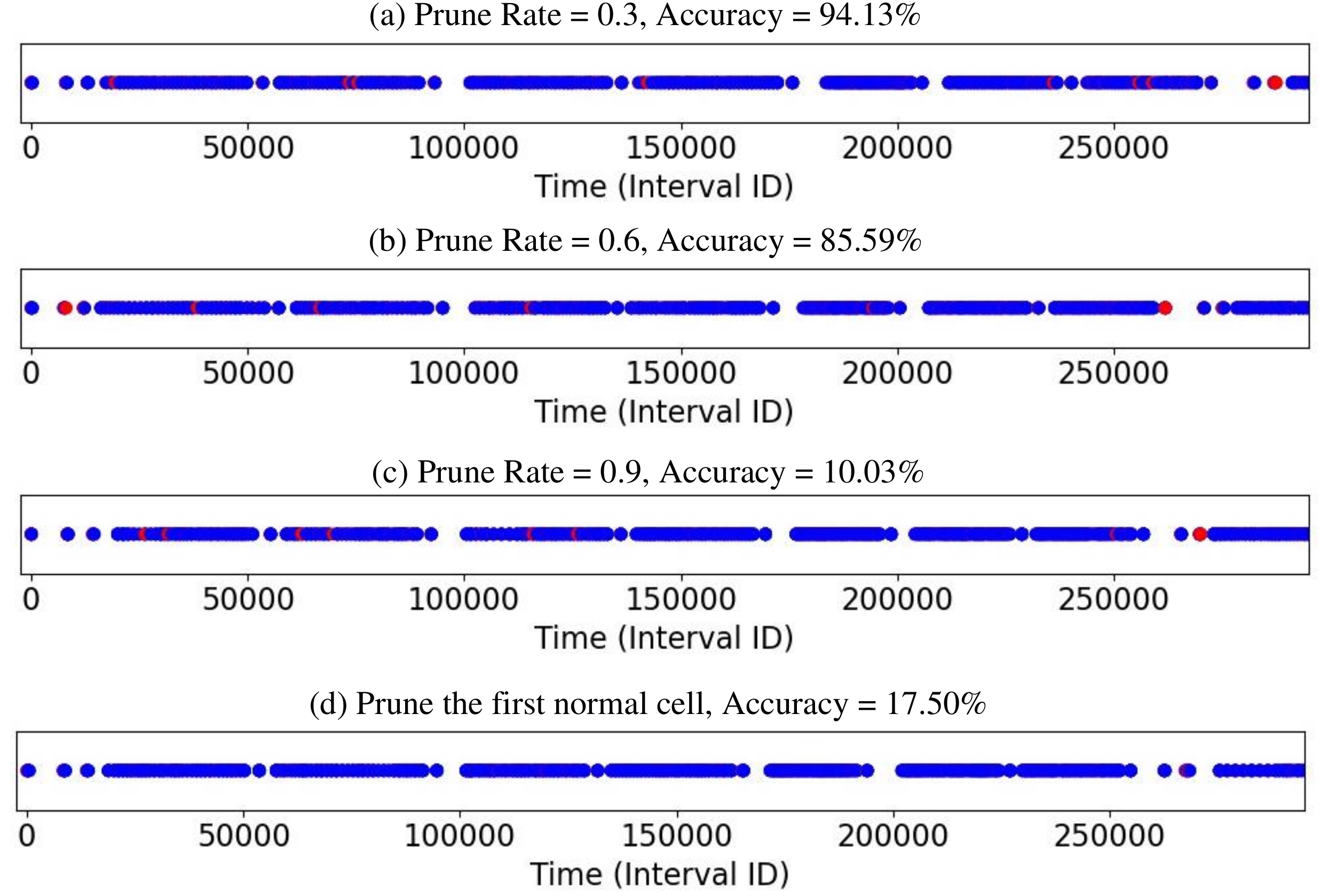}
  \vspace{-15pt}
  \caption{Side-channel traces of weighting pruned models.}
  \label{fig_prune}
%   \vspace{-15pt}
  \end{figure}

\noindent\textbf{Model transformation.}
Prior parameter-based solutions \cite{adi2018turning,zhang2020model,chen2021temporal} are proven to be vulnerable against model fine-tuning or image transformations \cite{chen2019refit,shafieinejad2019robustness,liu2020removing,guo2020hidden}. In contrast, our scheme is robust against these transformations as it only modifies the network architecture. 
First, we consider four types of fine-tuning operations evaluated in \cite{adi2018turning} (\emph{Fine-Tune Last Layer}, \emph{Fine-Tune All Layers}, \emph{Re-Train Last Layer}, \emph{Re-Train All Layers}). We verify that they do not corrupt our watermarks embedded to the model architecture.
Second, we consider model compression. Common pruning techniques set certain parameters to 0 to shrink the network size. The GEMM computations are still performed over pruned parameters, which give similar side-channel patterns. 
Figures \ref{fig_prune}(a)-(c) show the extraction trace of the first normal cell after the entire model is pruned with different rates (0.3, 0.6, 0.9) using $L_2$-norm. Figure \ref{fig_prune}(d) shows one case where we prune all the parameters in the first normal cell. We observe that a bigger pruning rate can decrease the length of the leakage window, as there are more zero weights to simplify the computation.
However, the pattern of the operations in the cell keeps unchanged, indicating the weight pruning cannot remove the embedded watermark.

% 1:89.34\%, 93.38\%, 78.71\%, 55.62\%
% 2:47.18\%, 70.87\%, 37.47\%, 64.89\%
% 3:44.5\%, 45.9\%, 28.61\%, 53.06\%
% 4:37.66\%

\noindent\textbf{Model obfuscation.}
An adversary may also obfuscate the inference execution to interfere with the verification results. (1) He can shuffle the orders of some operations which can be executed in parallel. However, since the selected stamp operations are in a path, they have high dependency and must be executed in the correct order. Hence, we can still identify the fixed operation sequence from the leakage trace of obfuscated models. 
(2) The adversary can add useless computations (e.g., matrix multiplications), operations or neurons to obfuscate the side-channel trace. Again, the critical stamp operations are still in the trace, and the owner is able to verify the ownership regardless of the extra operations. 
(3) The adversary may add useless cell windows to obfuscate the watermark verification.

\begin{figure}[t]
  \centering
    \includegraphics[width=0.9\linewidth]{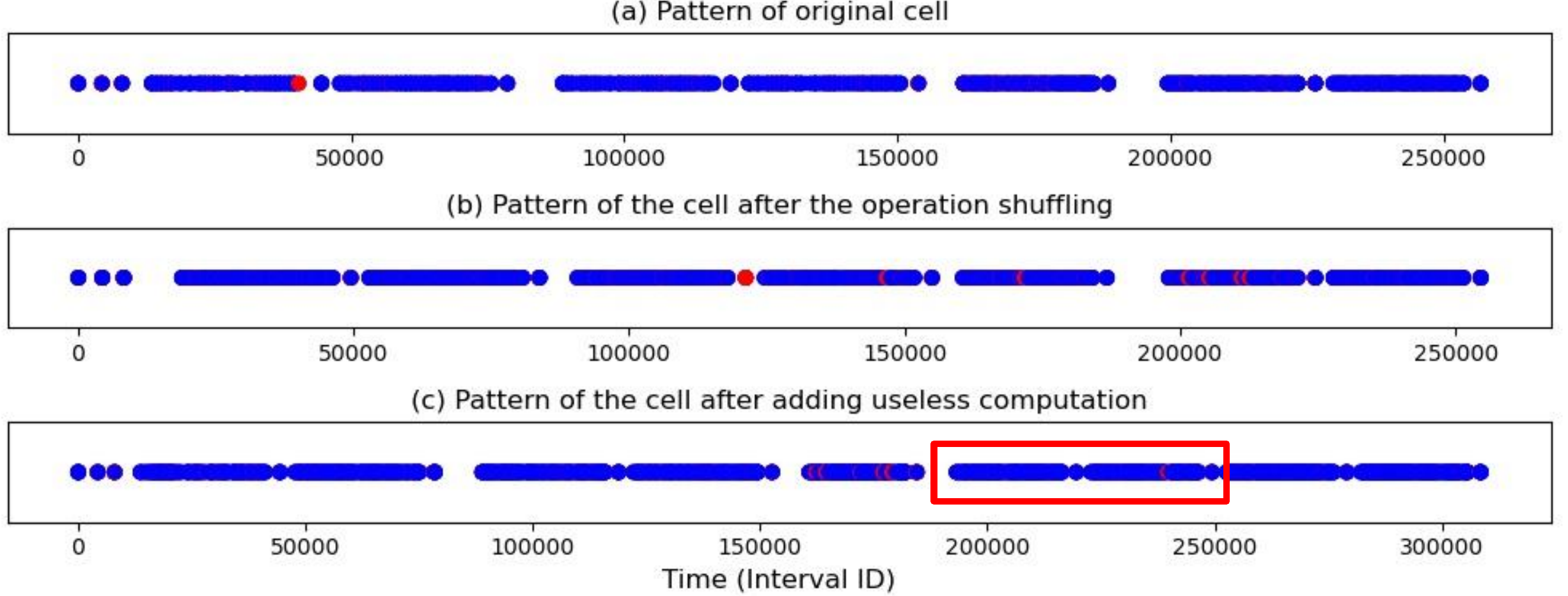}
  \caption{Traces of obfuscated models.}
  \label{fig_obfuse}
\vspace{-5pt}
\end{figure}

Figure \ref{fig_obfuse} illustrates the leakage pattern of the original cell as well as the cells after being obfuscated by above two techniques. Specifically, in Figure \ref{fig_obfuse}(b), the attacker shuffles the operation execution order, which first executes \circled{2}, \circled{4}, \circled{6} and \circled{8} and then runs the watermarked path. We can see that the watermark (i.e., fixed operations) can still be identified in the sequence. In Figure \ref{fig_obfuse}(c), the attacker adds an unused $3 \times 3$ separable convolution (red block) in the pipeline, which does not affect the watermark extraction, as the fixed sequence of stamp operations remains. In short, \emph{the stamp operations must be executed sequentially and cannot be removed in a lightweight manner.} This makes it difficult to remove the watermarks in the architecture. 

\begin{figure}[h]
  \centering
    \includegraphics[width=0.8\linewidth]{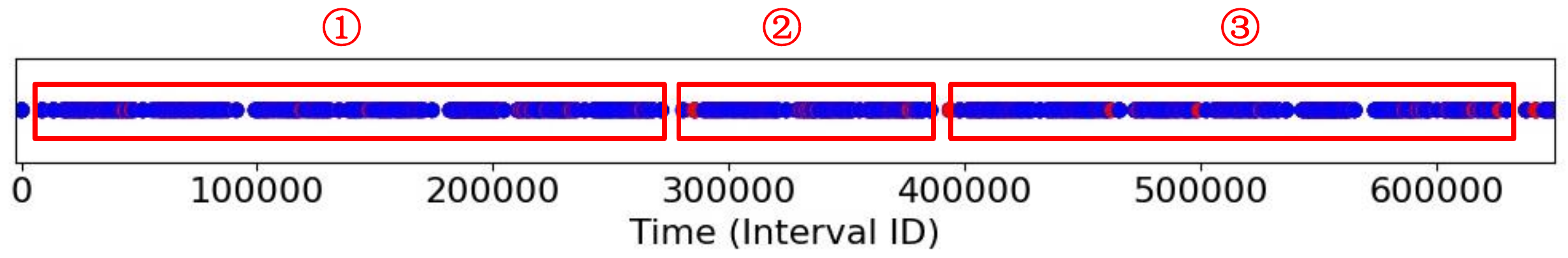}
  \caption{Influence of useless cell windows.}
  \label{fig_windows}
\end{figure}

Figure \ref{fig_windows} shows the influence of injected useless cell windows. In the side-channel trace, it contains three cell windows, where \circled{1} and \circled{3} are NAS cell windows and \circled{2} is the injected useless cell window. We just need to check if the monitored side-channel trace contains $N$ identical cells and identify if the watermark exists in the cells. Even there are other cells, we can also claim that this model is watermarked and then require for further arbitration.

\noindent\textbf{Structure pruning.}
We further consider the structured pruning, which can explicitly modify the model structure. This technique is indeed possible to remove our watermark embedded into the network architecture. However, it has two drawbacks: (1) since the watermarking key is secret, the adversary does not know which operation should be pruned; (2) Pruning the stamp operation can cause significant performance drop. To validate this, we random prune 1 to 4 stamp operations in the normal cell, and Table \ref{table_strprune} shows the prediction accuracy of pruned models on CIFAR10. For the case of pruning one stamp operation, we give four prediction accuracy values corresponding to four possible pruning scheme (pruning one operation from \circled{2}, \circled{3}, \circled{5} and \circled{7} in Fig. \ref{fig_norcell}). For models with more than one stamp operations pruned, we give the average accuracy of pruned models. We observe that even only pruning one stamp operation can lead to great accuracy drop (96.53\% to 55.62\%). Hence, removing the watermark with structured pruning is not practical.

\begin{table}[]
\centering
\resizebox{\linewidth}{!}{
\begin{tabular}{|c|c|c|c|c|c|}
\hline
\# of pruned stamp ops & 0 & 1 & 2 & 3 & 4 \\ \hline
Accuracy (\%) & 96.53 & \makecell[c]{89.34/93.38/ \\ 78.71/55.62 }& 54.89 &  44.52 & 37.66  \\ \hline
\end{tabular}}
\caption{Accuracy of structured pruned models on CIFAR10}\label{table_strprune}
\vspace{-10pt}
\end{table}

Note that an adversary can leverage some powerful methods (e.g., knowledge distillation \cite{ba2013deep,hinton2015distilling}) to fundamentally change the architecture of the target model and possibly erase the watermarks. However, this is not flagged as copyright violation, since the adversary needs to spend a quantity of effort and cost (computing resources, time, dataset) to obtain a new model. This model is significantly different from the original one, and is regarded as the adversary's legitimate property. %We do not consider such operations in our paper. 

\begin{figure}[h]
      \centering
        \includegraphics[width=0.9\linewidth]{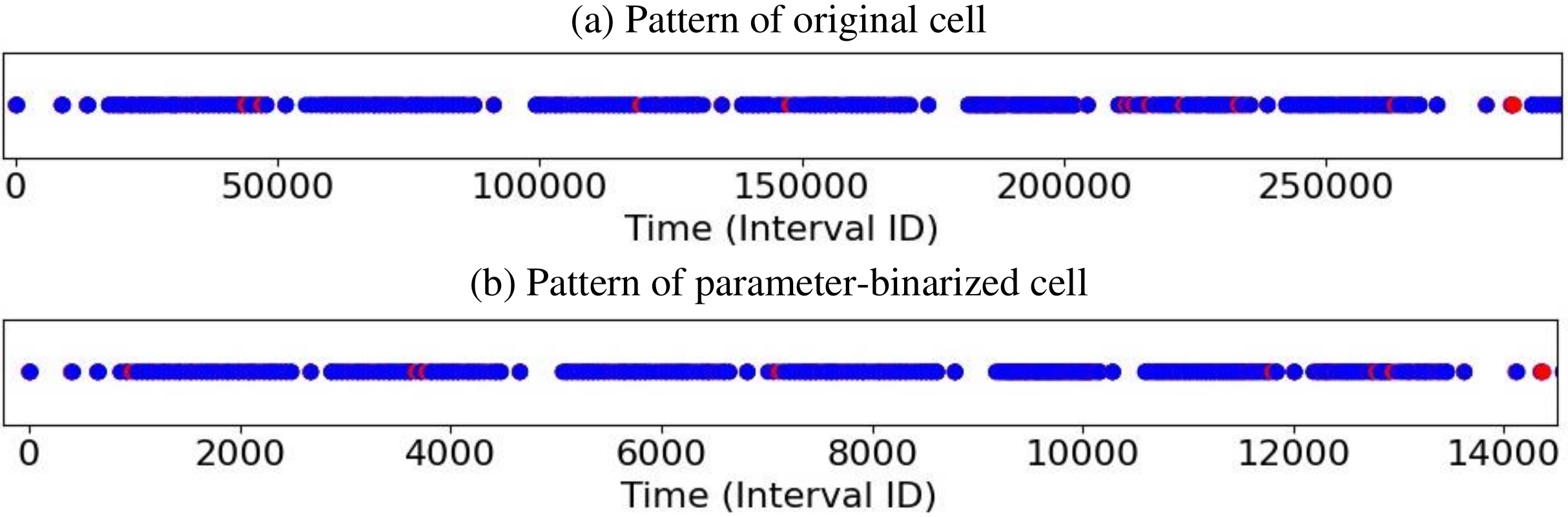}
      \caption{Influence of parameter binarization.}
      \label{fig_bnn}
    \end{figure}

\noindent\textbf{Parameter binarization.}
This technique \cite{courbariaux2016binarized} is used to accelerate the model execution by binarizing the model parameters. If corresponding Binary Neural Network (BNN) still adopts the BLAS library to accelerate the matrix multiplications, the side-channel leakage pattern keeps similar. Only the time interval between each monitored API access becomes shorter, as the parameter binarization would cause much faster model execution. Figure \ref{fig_bnn} shows the comparison of side-channel traces between the original NAS cell and binarized cell. We observe that although parameter binarization achieves about 20 times faster inference (2.8e5 vs. 1.4e4 intervals), the leakage trace still keeps the similar pattern. Hence, our scheme can still be applied to verify BNN models. If the BNN model adopts other acceleration libraries, we can also switch to monitor that library to perform similar analysis.  
    
%While both fine-tuning and model compression, in particular the parameter pruning, are considered in our experiments, other attacks like knowledge distillation that requires training from scratch and tremendous computation resource are out of the scope. 

% Even for the model distillation, the attacker needs to retrain on the whole dataset, it is out of our scope. 
% And noted that as $mk$ only marks certain designated edges and operations, the attacker only has possibility to totally remove the mark. 

% (2)remove entire nodes from the network, modifying the network architecture itself, while aiming to keep the accuracy of the initial larger network. Actually it is easy to hurt the accuracy. 

% Trying to remove the watermark with two attack methods, (1) fine-tune and (2) first pruning and then fine-tune. 

% The experiment one would assume the attacker knows internal details of the model and owns a small training set (20\%), then training the model for few epochs to figure out if fine-tune removes the mark. 

% The experiment two first prunes the model by randomly modifying or even removing some edges in the cell, then train it with small training set, checking if the watermark is removed and how is the prediction accuracy. Noted that as $mk$ only marks certain designated edges and operations, the attacker only has possibility of xxx\% (to compute) to totally remove the mark. 

%\subsection{Watermarking Verification}

\subsection{Uniqueness}\label{sec:uniqueness}

Given a watermarked model, we expect that benign users have very low probability to obtain the same architecture following the original NAS method. This is to guarantee small false positives of watermark verification. 

\begin{figure}[t]
    \centering
    \includegraphics[width=0.49\linewidth]{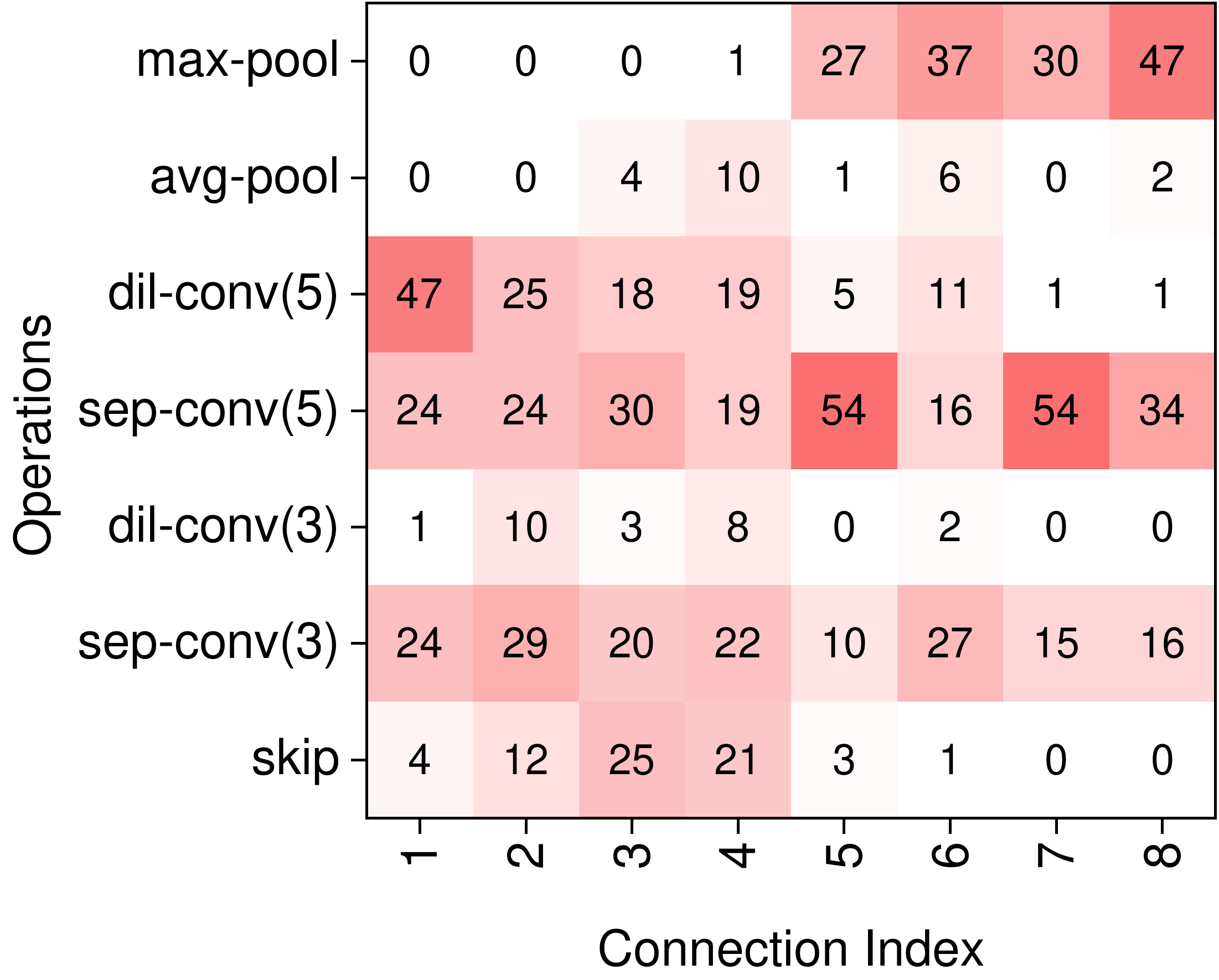}\label{fig_normalheat}
    \includegraphics[width=0.49\linewidth]{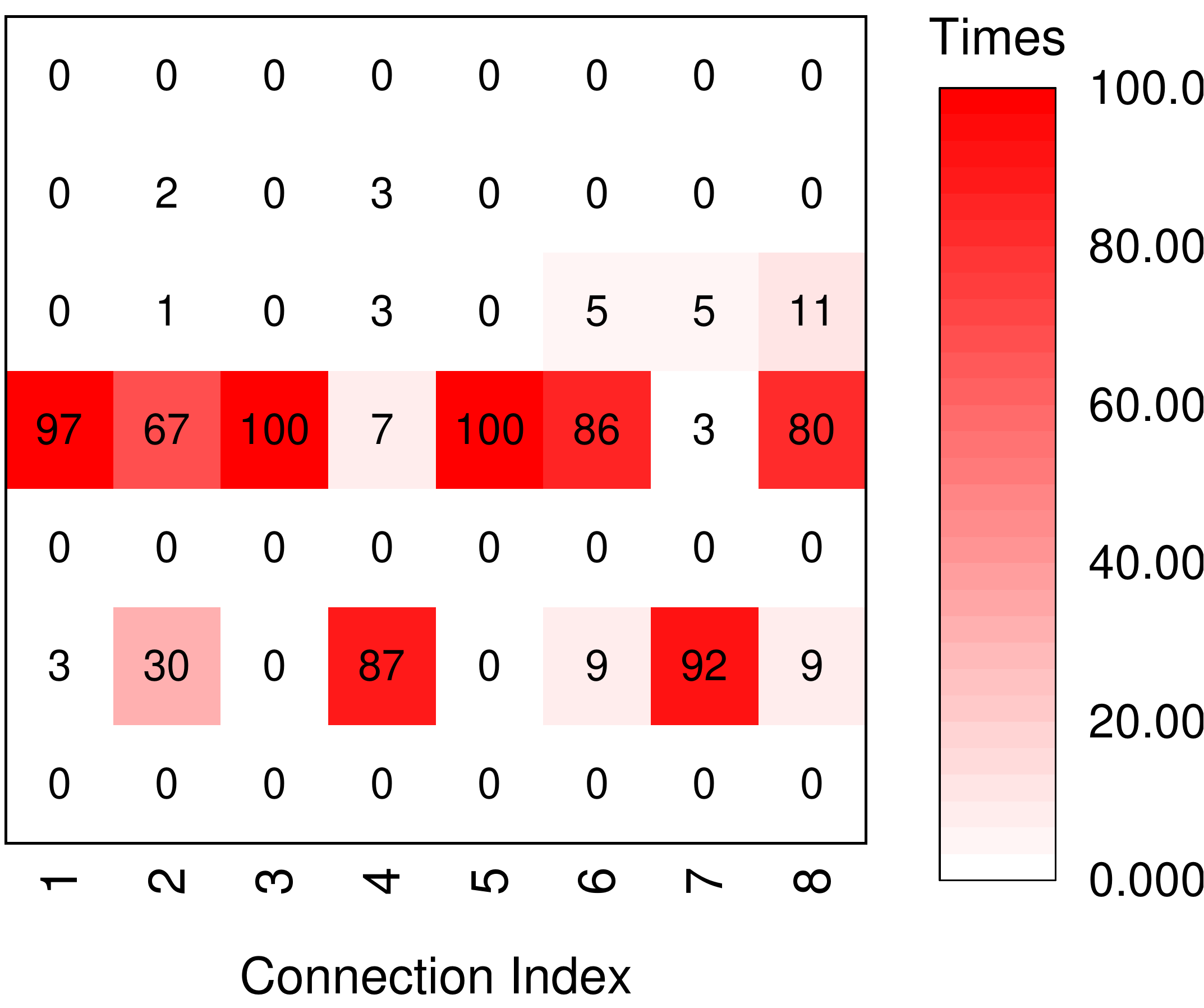}\label{fig_reduceheat}
    \vspace{-5pt}
    \caption{Operation distributions for a normal cell (left) and reduction cell (right). The connection index is the index of the connection edge in the NAS cell.} \label{fig_heat}
    \vspace{-5pt}
  \end{figure}

The theoretical analysis assumes each edge selects various operations with equal probability, and shows the collision rate is less than 0.03\% (see Appendix \ref{sec:proof}).
We further empirically evaluate the uniqueness of our watermarking scheme. Specifically, we repeat the GDAS method on CIFAR10 for 100 times with different random seeds to generate 100 architecture pairs for the normal and reduction cells. We find our stamps have no collision with these 100 normal models. Figure \ref{fig_heat} shows the distribution of the operations on eight connection edges in the two cells. We observe that most edges have some preferable operations, and there are some operations never attached to certain edges. This is more obvious in the architecture of the reduction cell. 
Such feature can help us to select more unique operation sequence as the marking key. Besides, the collision probability is decreased when the stamp size $n_s$ is larger. A stamp size of 4 with fixed edge-operation selection can already achieve strong uniqueness.

\section{Conclusion} \label{sec_conclusion}
In this paper, we propose a new direction for IP protection of DNNs. We show a carefully-crafted network architecture can be utilized as the ownership evidence, which exhibits stronger resilience against model transformations than previous solutions. We leverage Neural Architecture Search to produce the watermarked architecture, and cache side channels to extract the black-box models for ownership verification. Evaluations indicate our scheme can provide great effectiveness, usability, robustness, and uniqueness, making it a promising and practical option for IP protection of AI products. 

%practical watermarking scheme embedding to the model architecture instead of parameters. We utilize Neural Architecture Search to embed watermark and leverage cache side channels to verify the watermark, both of which have not been discussed before. Our comprehensive evaluations show that our approach has no noticeable impact on the model performance and is robust against practical attacks. Besides, our scheme first achieves watermarking a class of models with o ne marking key. 

\bibliography{references}
\bibliographystyle{IEEEtran}

% \newpage
\appendix

\subsection{Proof Sketch of Theorem 1}
\label{sec:proof}
\begin{proof}[Proof Sketch]
We prove that our algorithms (\textbf{WMGen}, \textbf{Mark}, \textbf{Verify}) form a qualified watermarking scheme for NAS models.

  \noindent\textbf{Effectiveness.} The property is guaranteed by Assumption \ref{ass:leakage}.

  \noindent\textbf{Usability.} Let $\mathbb{S}_{c_i,0}$, $\mathbb{S}_{c_i}$ be the architecture search spaces before and after restricting the stamp $k_i$ of $c_i$. $\mathfrak{A}_{c_i,0}$ and $\mathfrak{A}_{c_i}$  are the two architecture searched from $\mathbb{S}_{c_i,0}$ and $\mathbb{S}_{c_i}$, respectively. $f_{c_i,0}$ and $f_{c_i}$ are the corresponding models trained on the the same data distribution $\mathcal{D}$. From Assumption \ref{ass:stamp}, we have
  \begin{equation}
     Pr[f_{c_i,0}(x) = y|(x,y)\thicksim \mathcal{D}] - Pr[f_{c_i}(x) = y|(x,y)\thicksim \mathcal{D}]\leq  \frac{\epsilon}{N}.
  \end{equation}
 
Let $f_0, f$ are the DNN models that are learned before and after restricting their architecture search spaces by a watermark. One can easily use the mathematical induction to prove the usability of our watermarking scheme, i.e.,
\begin{equation}
Pr[f_0(x) = y|(x,y)\thicksim \mathcal{D}] - Pr[f(x) = y|(x,y)\thicksim \mathcal{D}] \leq \epsilon.
\end{equation}

\noindent\textbf{Robustness.} We classify the model modification attacks into two categories. The first approach is to only change the parameters of $f$ using existing techniques such as fine-tuning and model compression. Since the architecture is preserved, the stamps of all cells are also preserved. According to Assumption \ref{ass:leakage}, the idea analyzer can extract the stamps and verify the ownership of the modified models.

 The other category of attacks modifies the architecture of the model. Since the marking key (watermark) is secret, the adversary can uniformly modify the operation of an edge or delete an edge in a cell. The probability that the adversary can successfully modify one edge/operation of a stamp is not larger than $\frac{n_s}{|c_i|}$, where $|c_i|$ is the number of connected edges in $c_i$. Thus, the expected value of the total number of modification is $\delta \times \frac{N\times n_s}{\sum_{i=1}^N|c_i|}^{\tau\times N_s}$. However, since the adversary cannot access the proxy and task datasets, he cannot obtain new models with competitive performance by retraining the modified architectures.

%   \noindent\textbf{Uniqueness.}
%   Without loss of generality, we assume the NAS algorithm can search the same architecture if the search spaces of all cells are the same. Thus, the uniqueness of the watermarked model is decided by the probability that the adversary can identify the same search spaces. Because the marking key is secret, the adversary has to guess the edges and the corresponding operations of each stamp if he wants to identify the same search spaces (the similarity ratio $\tau$ is 100\% here). The probability that he can identify the same search space of a cell is much smaller than $(\frac{1}{\mathcal{|O|}})^{n_s}$. Therefore, the uniqueness property of our watermarking scheme is proved if $\delta < (\frac{1}{\mathcal{|O|}})^{n_s\times N}$. As we empirically evaluated in Section \ref{sec:uniqueness}, the watermarked models are unique and the probability of model collision is very low.
  
\noindent\textbf{Uniqueness.}  
Given a watermarked model, we expect that benign users have a very low probability to obtain the same architecture following the original NAS method. Without loss of generality, we assume the NAS algorithm can search the same architecture if the search spaces of all cells are the same. Thus, the uniqueness of the watermarked model is decided by the probability that the adversary can identify the same search spaces. Because the marking key is secret, the adversary has to guess the edges and the corresponding operations of each stamp if he wants to identify the same search spaces.
% This is to guarantee small false positives of watermark verification. 
Assume the selection of candidate operations is independent and identically distributed, the probability that an operation is chosen on an edge is $\frac{1}{|\mathcal{O}|}$. For a DNN model that contains $\mathcal{B}$ computation nodes, there are $2\mathcal{B}$ connection edges, from which we select $n_s$ causal edges. There are $\tbinom{2\mathcal{B}}{n_s}$ combinations. Hence, the probability of the stamp collision in a cell can be computed as $\tbinom{2\mathcal{B}}{n_s} \times (\frac{1}{|\mathcal{O}|})^{n_s}$. In our experiment configurations, the collision rate is smaller than 1.7\%. Considering both the normal and reduction cells, the collision rate is smaller than $(1.7\%)^2 \approx 0.03\%$, which can be neglected.

%  \noindent\textbf{Generalization.} This property can be directly proved because our watermarking scheme is task-agnostic and dataset-agnostic.
  \end{proof}
  
%Adi et al. \cite{adi2018turning} adopts an trusted third party and a commitment scheme to ensure the unforgeability property. One can easily adopt this technique into our watermarking scheme and make the embedded watermarks unforgeable.

\subsection{Get Path from Cell Supernet} \label{sec_path}
Algorithm \ref{alg_getpath} illustrates how to extract consecutive paths from the cell \emph{supernet} $\mathcal{G}$. The operation $\{\emph{set}\} \circ \mathcal{N}_i$ appends the node $\mathcal{N}_i$ to each element in the \emph{set}, generating a set $P_i$ of possible paths from the cell inputs to node $\mathcal{N}_i$. Specifically, $P_\mathcal{B}$ contains all the candidate paths in the cell supernet $\mathcal{G}$. Given the number of fixed stamp edges $n_s$ , our goal is to identify a path of length $n_s$ from $\mathcal{G}$. Note that the longest consecutive path in $\mathcal{G}$ contains $\mathcal{B}$ edges, so that it has $1 \leq n_s \leq \mathcal{B}$.
For each candidate path $p$ in $P_\mathcal{B}$, if its length is larger than $n_s$, we would extract all the subpaths with length $n_s$ from it (\texttt{GetSubPath}), and save them to $S_e$.

%\begin{minipage}{0.48\textwidth}\small
\begin{algorithm}[h]\small
	\caption{\texttt{GetPath} from Cell Supernet} \label{alg_getpath}
	\KwIn{cell supernet $\mathcal{G}$, \# of fixed edges $n_s$}
	\KwOut{set $S_e$ of all the possible paths with length $n_s$}
	$S_e = \{ \}$, \\ $P_1 = \{a, b\} \circ \mathcal{N}_1$  \\
	\For{i from 2 to $\mathcal{B}$}{
	    $P_i = (P_{i-1} \cup ... \cup P_1 \cup \{a\} \cup \{b\}) \circ \mathcal{N}_i$ \\
	}
	\For{$p$ in $P_\mathcal{B}$}{
	    \If{$|p| \geq n_s$}{
	        $S_e = S_e \cup \texttt{GetSubPath}(p, n_s)$ \\
	    }
	}
	\Return $S_e$
\end{algorithm}
%\end{minipage}
%\hfill
%\begin{minipage}{0.48\textwidth}\small
\iffalse
\begin{algorithm}[h]
	\caption{\emph{GEMM} in OpenBLAS} \label{alg_gemm}
	\KwIn{matrice A, B, C; scalars $\alpha$, $\beta$}
	%classifier setting ($L$, $P$) }
	\KwOut{$C=\alpha A \times B + \beta C$}
	\For(\tcp*[h]{Loop 1}){ j in (0:R:n)}{
	    \For(\tcp*[h]{Loop 2}){ l in (0:Q:k)}{
	        call \emph{itcopy} \\ 
	        \For(\tcp*[h]{Loop 4}){ jj in (j:3UNROLL:j+R)}{
	            call \emph{oncopy} \\
	            call \emph{kernel} \\
	        }
	        \For(\tcp*[h]{Loop 3}){ i in (P:P:m)}{
	            call \emph{itcopy} \\
	            call \emph{kernel} \\
	        }
	    }
	}  
\end{algorithm}
\fi
%\end{minipage}

\subsection{Details about GEMM in OpenBLAS} \label{app_gemm}
\begin{figure}[h]
  \centering
  \includegraphics[width=0.6\linewidth]{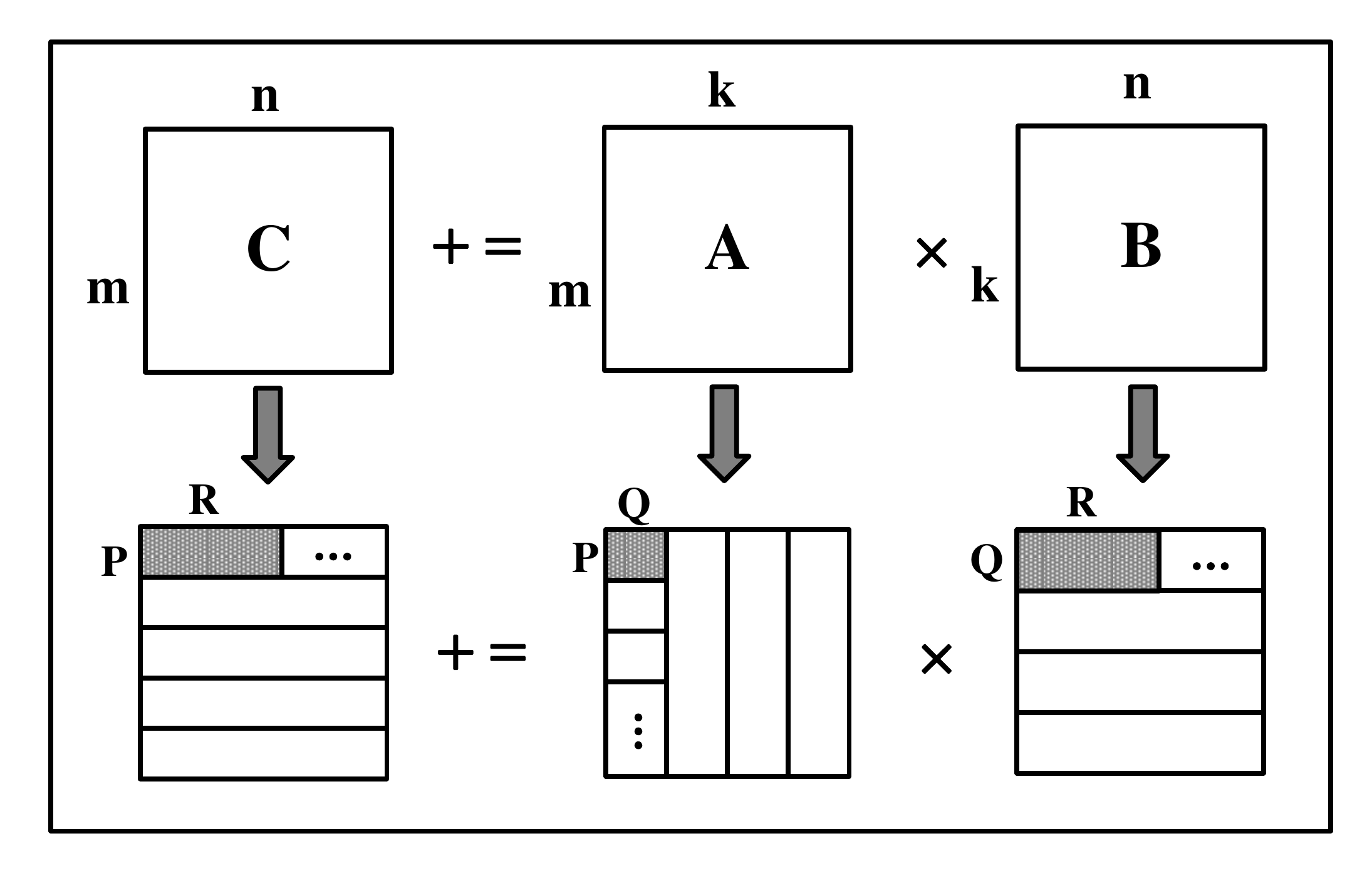}
  \vspace{-10pt}
  \caption{The procedure of GEMM.}
  \label{fig_gemm}
  \vspace{-10pt}
\end{figure}

BLAS realizes the matrix multiplication with the function \emph{gemm}. This function computes $C=\alpha A \times B + \beta C$, where A is an $m \times k$ matrix, B is a $k \times n$ matrix, C is an $m \times n$ matrix, and both $\alpha$ and $\beta$ are scalars. OpenBLAS adopts Goto's algorithm \cite{goto2008anatomy} to accelerate the multiplication using modern cache hierarchies. This algorithm divides a matrix into small blocks (with constant parameters P, Q, R), as shown in Figure \ref{fig_gemm}.
The matrix A is partitioned into $P \times Q$ blocks and B is partitioned into $Q \times R$ blocks, which can be fit into the L2 and L3 caches, respectively. The multiplication of such two blocks generates a $P \times R$ block in the matrix C.
Algorithm \ref{alg_gemm} shows the process of \emph{gemm} that contains 4 loops controlled by the matrix size $(m, n, k)$. Functions \emph{itcopy} and \emph{oncopy} are used to allocate data and functions. \emph{kernel} runs the actual computation. Note that the partition of $m$ contains two loops, $loop_3$ and $loop_4$, where $loop_4$ is used to process the multiplication of the first $P \times Q$ block and the chosen $Q \times R$ block.
For different cache sizes, OpenBLAS selects different values of P, Q and R to achieve the optimal performance.
% In our testbed (Section \ref{sec_setup}), the function \emph{sgemm\_nn} used by DNN adopts $P=320$, $Q=320$, $R=104512$ and $UNROLL=4$. Since $R$ is larger than most input $n$, at least in the scenarios of NAS, we assume $loop_1$ is only performed once.

\begin{algorithm}[t]\small
	\caption{\emph{GEMM} in OpenBLAS} \label{alg_gemm}
	\KwIn{matrice A, B, C; scalars $\alpha$, $\beta$}
	%classifier setting ($L$, $P$) }
	\KwOut{$C=\alpha A \times B + \beta C$}
	\For(\tcp*[h]{Loop 1}){ j in (0:R:n)}{
	    \For(\tcp*[h]{Loop 2}){ l in (0:Q:k)}{
	        call \emph{itcopy} \\ 
	        \For(\tcp*[h]{Loop 4}){ jj in (j:3UNROLL:j+R)}{
	            call \emph{oncopy} \\
	            call \emph{kernel} \\
	        }
	        \For(\tcp*[h]{Loop 3}){ i in (P:P:m)}{
	            call \emph{itcopy} \\
	            call \emph{kernel} \\
	        }
	    }
	}  
\end{algorithm}

\subsection{Details about the NAS Algorithms} \label{app_nas}
\subsubsection{Architecture Search}
We adopt GDAS \cite{dong2019searching} to search for the optimal CNN architectures on CIFAR10.
% To carry out architecture search, we randomly split the official training data into two groups, one is training set and the other is validation set, with each group containing 25K images. 
We set the number of initial channels in first convolution layer as 16, the number of the computation nodes in a cell as 4 and the number of normal cells in a block as 2. Then we train the model for 240 epochs. The setting of the optimizer and learning rate schedule is the same as that in \cite{dong2019searching}. The search process on CIFAR10 takes about five hours with a single NVIDIA Tesla V100 GPU. 

% \noindent\textbf{PTB.}
% We adopt DARTS \cite{liu2018darts} to search for the optimal RNN architecture on PTB. Both the embedding and hidden sizes are set as 300, and the network is trained for 50 epochs using SGD optimization. We set the learning rate as 20, the batch size as 256, BPTT length as 35, and the weight decay as $5 \times 10^{-7}$. Other setting of the optimization of the architecture is also the same as \cite{liu2018darts}. The search process takes 6 hours on a single GPU.

\subsubsection{Model Retraining}
After obtaining the searched cells, we construct the CNNs for CIFAR and ImageNet. For the CIFAR, we form a CNN with  33 initial channels. We set number of computation nodes in a cell as 4 and the number of normal cells in a block as 6. Then we train the network for 300 epochs on the dataset (both CIFAR10 and CIFAR100), with a learning rate reducing from 0.025 to 0 with the cosine schedule. The preprocessing and data augmentation is the same as \cite{dong2019searching}. The training process takes about 11 GPU hours. For the CNN on ImageNet, we set the initial channel size as 52, and the number of normal cells in a block as 4. The network is trained with 250 epochs using the SGD optimization and the batch size is 128. The learning rate is initialized as 0.1, and is reduced by 0.97 after each epoch. The training process takes 12 days on a single GPU.

% \noindent\textbf{PTB}
% A RNN with the searched recurrent cell is trained on PTB with the SGD optimization and the batch size of 64 until the convergence. Both the embedding and hidden sizes are set as 850. The learning rate is set as 20 and the weight decay is $8 \times 10^{-7}$. The training process takes 3 days on a single GPU. 

\subsection{Monitored Functions in Pytorch and OpenBLAS} \label{app_monitor}
\begin{figure}[h]
  \centering
  \includegraphics[width=\linewidth]{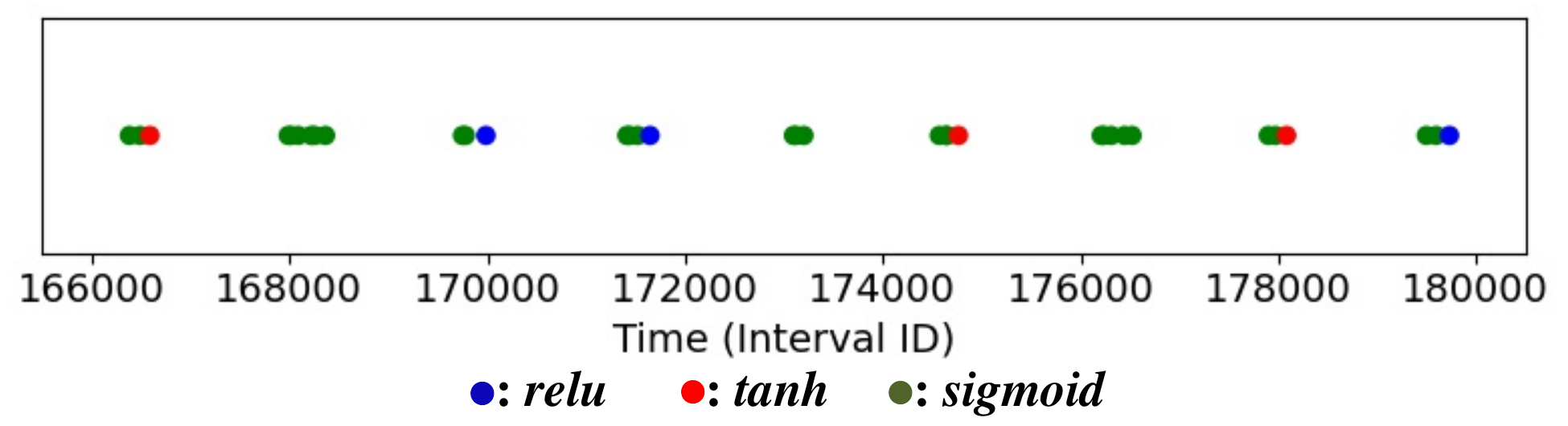}
  \vspace{-10pt}
  \caption{The side-channel trace of a recurrent cell.}
  \label{fig_rnn}
  \vspace{-10pt}
\end{figure}

Table \ref{table_rnn} gives the monitored code lines in the latest Pytorch 1.8.0 and OpenBLAS 0.3.15. To identify computationally intensive operations (i.e., convolutions), we need to monitor accesses to functions \emph{itcopy} and \emph{oncopy}. To protect other DNN components like activations, we turn to monitor corresponding activation APIs in the library. Figure \ref{fig_rnn} shows an example side-channel leakage trace of activation functions monitored from a NAS cell. We observe that the trace contains 9 separate clusters, each of which represents the existence of activation functions in a DNN model layer.

\begin{table}[htb]
\centering
\resizebox{\linewidth}{!}{
\begin{tabular}{|l|l|l|}
\hline
Library & Functions & Code Line  \\ \hline
\multirow{2}{*}{OpenBLAS} & Itcopy & kernel/generic/gemm\_tcopy\_8.c:78 \\
\cline{2-3}
 & Oncopy & kernel/x86\_64/sgemm\_ncopy\_4\_skylakex.c:57 \\
\hline
\multirow{5}{*}{Pytorch} & Relu & aten/src/ATen/Functions.cpp:6332 \\
\cline{2-3}
 & Tanh & aten/src/ATen/native/UnaryOps.cpp:452 \\
\cline{2-3}
 & Sigmoid & aten/src/ATen/native/UnaryOps.cpp:389 \\
\cline{2-3}
 & Avgpool & aten/src/ATen/native/AdaptiveAveragePooling.cpp:325 \\
\cline{2-3}
 & Maxpool & aten/src/ATen/native/Pooling.cpp:47 \\
\hline
\end{tabular}}
\caption{Monitored code lines in OpenBLAS and Pytorch.}
\label{table_rnn}
\vspace{-10pt}
\end{table}

% \subsection{Whole side channel leakage trace} \label{app_trace}
% Figure \ref{fig_wholetrace} shows the whole side channel leakage trace of the tested NAS model in our end-to-end watermarking process. While the nodes representing the function accesses are stacked up, we can still identity the first block from interval 0 to around $2 \times 1e6$, where there are more accesses to \emph{itcopy} (blue nodes). For the following two blocks, since the number of channels increases, the length of leakage windows also increases. 

% \begin{figure}[H]
%   \centering
%   \includegraphics[width=\linewidth]{./fig/fig_wholetrace.pdf}
%   \caption{Whole leakage trace of the NAS model.}
%   \label{fig_wholetrace}
%   \vspace{-10pt}
% \end{figure}

\vfill

\end{document}